\documentclass[prx,twocolumn,superscriptaddress]{revtex4-1}

\usepackage{amssymb}
\usepackage{amsfonts}
\usepackage{graphics,graphicx,epsfig,bm,amsmath,amsthm,amssymb}
\usepackage{bm}
\usepackage{bbm}
\usepackage{longtable}	
\usepackage{multirow}
\usepackage{array}
\usepackage{color}
\usepackage{ulem,comment}
\usepackage[usenames,dvipsnames]{xcolor}
\usepackage{amsthm}
\usepackage{appendix}

\usepackage{float}
\usepackage[colorlinks=true,
linkcolor=blue,citecolor=blue,
pdfauthor={ },
pdftitle={ },
pdfsubject={ },
pdfkeywords={ }]{hyperref}

\newcommand{\rC}{r_\text{\tiny C}}
\newcommand{\D}{\operatorname{d}\!}
\graphicspath{{../Figures/}}

\begin{document}

\title{Room temperature test of the Continuous Spontaneous Localization model using a levitated micro-oscillator}

\author{Di Zheng}
\affiliation{National Laboratory of Solid State Microstructures and Department of Physics, Nanjing University, Nanjing 210093, China}

\author{Yingchun Leng}
\affiliation{National Laboratory of Solid State Microstructures and Department of Physics, Nanjing University, Nanjing 210093, China}

\author{Xi Kong}
\affiliation{National Laboratory of Solid State Microstructures and Department of Physics, Nanjing University, Nanjing 210093, China}


\author{Rui Li}

\affiliation{Hefei National Laboratory for Physical Sciences at the Microscale and Department of Modern Physics, University of Science and Technology of China, Hefei 230026, China}
\affiliation{CAS Key Laboratory of Microscale Magnetic Resonance, University of Science and Technology of China, Hefei 230026, China}
\affiliation{Synergetic Innovation Center of Quantum Information and Quantum Physics, University of Science and Technology of China, Hefei 230026, China}

\author{Zizhe Wang}

\affiliation{Hefei National Laboratory for Physical Sciences at the Microscale and Department of Modern Physics, University of Science and Technology of China, Hefei 230026, China}
\affiliation{CAS Key Laboratory of Microscale Magnetic Resonance, University of Science and Technology of China, Hefei 230026, China}
\affiliation{Synergetic Innovation Center of Quantum Information and Quantum Physics, University of Science and Technology of China, Hefei 230026, China}

\author{Xiaohui Luo}
\affiliation{National Laboratory of Solid State Microstructures and Department of Physics, Nanjing University, Nanjing 210093, China}

\author{Jie Zhao}
\affiliation{Hefei National Laboratory for Physical Sciences at the Microscale and Department of Modern Physics, University of Science and Technology of China, Hefei 230026, China}
\affiliation{CAS Key Laboratory of Microscale Magnetic Resonance, University of Science and Technology of China, Hefei 230026, China}
\affiliation{Synergetic Innovation Center of Quantum Information and Quantum Physics, University of Science and Technology of China, Hefei 230026, China}

\author{Chang-Kui Duan}
\affiliation{Hefei National Laboratory for Physical Sciences at the Microscale and Department of Modern Physics, University of Science and Technology of China, Hefei 230026, China}
\affiliation{CAS Key Laboratory of Microscale Magnetic Resonance, University of Science and Technology of China, Hefei 230026, China}
\affiliation{Synergetic Innovation Center of Quantum Information and Quantum Physics, University of Science and Technology of China, Hefei 230026, China}

\author{Pu Huang}
\email{hp@nju.edu.cn}
\affiliation{National Laboratory of Solid State Microstructures and Department of Physics, Nanjing University, Nanjing 210093, China}

\author{Jiangfeng Du}
\email{djf@ustc.edu.cn}
\affiliation{Hefei National Laboratory for Physical Sciences at the Microscale and Department of Modern Physics, University of Science and Technology of China, Hefei 230026, China}
\affiliation{CAS Key Laboratory of Microscale Magnetic Resonance, University of Science and Technology of China, Hefei 230026, China}
\affiliation{Synergetic Innovation Center of Quantum Information and Quantum Physics, University of Science and Technology of China, Hefei 230026, China}

\author{Matteo Carlesso}
\affiliation{Department of Physics, University of Trieste, Strada Costiera 11, 34151 Trieste, Italy}
\affiliation{Istituto Nazionale di Fisica Nucleare, Trieste Section, Via Valerio 2, 34127 Trieste, Italy}

\author{Angelo Bassi}
\affiliation{Department of Physics, University of Trieste, Strada Costiera 11, 34151 Trieste, Italy}
\affiliation{Istituto Nazionale di Fisica Nucleare, Trieste Section, Via Valerio 2, 34127 Trieste, Italy}


\begin{abstract}
{The} Continuous Spontaneous Localization (CSL) model predicts a tiny break of energy conservation via a weak stochastic force acting on physical systems, {which triggers the collapse of the wave function}.
Mechanical oscillators are a natural way to test such {a} force; in particular levitated micro-mechanical oscillator has been recently proposed to be an
ideal system. We {report a proof-of-principle experiment with} a micro-oscillator generated by a micro-sphere diamagnetically levitated in a magneto-gravitational trap under high vacuum. Due to the ultra-low mechanical dissipation, the
oscillator provides a new upper bound on the CSL collapse rate, {which gives} an improvement of two orders of magnitude over the previous bounds in the same frequency range, and partially reaches the enhanced collapse
rate suggested by Adler.
Although being performed at room temperature, our experiment has already exhibits advantages over those operating at low temperatures previously reported.
Our results experimentally show the potential of
magneto-gravitational levitated mechanical oscillator as a promising method for testing collapse model. {Further improvements in} cryogenic experiments are discussed.

\end{abstract}
\maketitle

\section{Introduction}

The perceived absence of macroscopic quantum superposition  attracts {the} physicists' interests {since}  the birth of quantum mechanics.  Different interpretations and reformulations {of quantum mechanics \cite{Bohm1,Bohm2,Durr,manyworlds1,manyworlds2, wallace, dariano} have been} proposed to comprehensively handle such an issue, however,  {most} of them {do not} provide direct
experimental testability.

{A phenomenological and experimentally verifiable \cite{Pearle1989,GRW,Ghirardi1990,Bassi2003,Bassi2013,Diosi2014,Nimmrichter2014,bahrami2014} approach is proposed by collapse models \cite{Leggett}. They}
introduce nonlinear and stochastic {terms in} the Schr\"{o}dinger equation, {which induce a spontaneous collapse of the wave function. Such a collapse is the stronger the larger the system.} {The origin of the noise remains an open question, and often in the literature it has been linked to}
gravity~\cite{Diosi1989,Penrose1996,Penrose}. In {this paper we focus on} the Continuous Spontaneous Localization (CSL) model, {one of the most studied in the literature}.

{CSL is characterized by two phenomenological parameters: the collapse rate $\lambda$ and its correlation length $\rC$. The latter can be also understood as the minimum superposition separation necessary to trigger a collapse. The theoretically suggested values for these parameters are $\lambda\simeq10^{-16}\,$s$^{-1}$ and $\rC=10^{-7}\,$m by Ghirardi \textit{et al.} \cite{GRW,Ghirardi1990}; while larger value are considered by Adler: $\lambda\simeq10^{-8\pm2}\,$s$^{-1}$ at $\rC=10^{-7}\,$m and $\lambda\simeq10^{-6\pm2}\,$s$^{-1}$ at $\rC=10^{-7}\,$m \cite{Adler2007}.}

{The CSL} modifications to standard quantum mechanics cause {primarily a loss of} coherence, {which} has been studied using {cold atoms \cite{Kovachy2015}}, molecular interferometers~\cite{Coldmolecular1999,Coldmolecular2011,Coldmolecular2012,Eibenberger2013,Interf1,Interf2}, {and} phonon motion in diamonds \cite{Diamondsexp,Diamonds}. {These compose the first class of experiments, namely the interferometric ones.} {Collapse models also predict a} stochastic force noise acting on the system, {whether it is in a quantum or classical state. This opens the way to non-interferometric experiments} \cite{Bassi2003,Bassi2013,Diosi2014},
{which has already constrained significantly the parameter space of the CSL model. To this class belong experiments with} ultra-cold atoms~\cite{ultra-cold atom,ultra-cold atom2}, bulk solid matter \cite{residual_heating,Adler2018,Phonos2}, planetary temperature observation \cite{Planetary,Planets,Planets2},  spontaneous X-ray emission~\cite{x-raycase,X-ray,X-ray2}, {and optomechanical systems \cite{AURIGA,LIGO1,LISA1,LISA2,gravitation results1,gravitation results2,Vinante2016,Vinante2017,Pontin2019}. Many more proposals were analyzed \cite{Collett2003,Rotational1,levitation3,MAQRO,McMillen2017,rotschrinski,multilayer,mishra2018}.}

{Non-interferometric experiments test the collapse mechanism at different frequencies, ranging from mHz \cite{LISA1,LISA2} to $10^{19}$ Hz \cite{X-ray}. Since the CSL noise is originally assumed to be white, the bound on the collapse parameters is independent from the frequency at which the collapse mechanism is probed. However, this is not the case for colored extensions of the model, where the noise is not white anymore and is typically characterized by a frequency cut-off \cite{Adler2007,colored1,colored2,colored3}.   }

{Recently, several studies \cite{Interf1,ultra-cold atom2,Colored-opto,Adler2018} were performed in this direction, indicating the urgency of testing the CSL noise at different frequencies to probe its spectrum. } {Opto-mechanics provides the optimal platform for this scope, since frequencies range from sub-mHz to kHz or even higher \cite{opto-rev}.}
{Among them, the gravitational wave detectors AURIGA, Advanced LIGO and LISA Pathfinder, due to their large test mass, succeeded in setting strong bounds on collapse parameters \cite{gravitation results1,gravitation
results2,Rotational1,gravitation results3} at}
{frequencies} less than $1$\,kHz \cite{AURIGA}, tens {of} Hz
\cite{LIGO1}, {and} sub-Hz \cite{LISA1,LISA2} {respectively.} {Among them, LISA Pathfinder gives the most strongest} upper bound {on $\lambda$} \cite{LISA2,Rotational1}. {Also} micro-scale
solid-state force sensors such as nano-cantilevers provided precise {testing} of the collapse {noise} \cite{Vinante2016,Vinante2017} at frequencies above kHz. {In this case, the relative large damping rates are balanced by operating at millikelvin temperatures.}

Levitated micro- or nano-mechanical oscillators are ideal for {potentially testing} {collapse models {due to their} low damping rates. Although they}
{recently} attracted considerable theoretical interest \cite{levitation1,levitation2,levitation3,levitation4,gravitation results3}, an experimental demonstration of their ability for such a purpose {has not been performed yet}.

Here, we report a {first proof-of-priciple }test of CSL based on a magnetically levitated micro-mechanical oscillator at room temperature. The levitation is realized {with} a specially designed magneto-gravitational trap
where a test particle {of mass of 4.7\,pg ($\sim2.8\times 10^{12}$\,amu)} is stably levitated for {some days in high vacuum}. {We observed a damping
rate
$\gamma/2\pi$ of the order of $30\,\mu$Hz at a resonant frequency of the order of 10\,Hz. This underlines the noiseless character of magneto-gravitational traps, that can actually provide a sensitive instrument for collapse model testing.} {As we will discuss in the following paragraphs, for} $\rC = 10^{-7}\,$m, we estimate the upper bound {$\lambda = 10^{-6.4}\,$s$^{-1}$ }{on the collapse rate}  at
 $95 \%$ confidence level{, excluding part of the range of values of the CSL parameters suggested by Adler~\cite{Adler2007}}. {This is} a significant improvement {with respect to the bound obtained from} the gravitational-wave detector Advanced LIGO {which operates} at the same frequency range~\cite{LIGO1,gravitation results1}{ and proves that magneto-gravitational levitation is a strong competitive platform for testing the limits of quantum mechanics}.

\section{Theoretical model}
According to the {mass-proportional} CSL model~\cite{Bassi2003}, {the collapse of the} wave-function {leads to a spontaneous diffusion }process{, which is} described by {the Lindblad term} \cite{Nimmrichter2014, bahrami2014, Diosi2014}:
\begin{equation}
\mathcal{L}_{\rm CSL}[\hat{\rho}_s(t)]=-\eta_i [\hat{x}_{i},[\hat{x}_{i},\hat{\rho}_s (t)]],
\end{equation}
where $\hat\rho_{s}(t)$ is the density operator {describing the }center {of mass} motion, $i=x,y,z$ labels the  direction {of motion}, and
\begin{equation}
\eta_i=\frac{\rC^{3}\lambda}{\pi^{\frac{3}{2}}m_{0}^{2}}\int d^{3}\mathbf{k}\,|\tilde \mu_{s}(\mathbf{k})|^{2}k_{i}^{2}e^{-k^{2}\rC^{2}}
\end{equation}
is {the CSL diffusion constant, which depends on the geometry of the object}. {Here}, $m_{0}$ is the atomic mass unit, $k = |\mathbf{k}|$ with $\mathbf{k} = (k_x,k_y,k_z)$, and $\tilde \mu_{s}(\mathbf{k})$ is the Fourier {transform} of the mass density
$\mu_{s}( \mathbf r)$, i.e. $\tilde \mu_{s}(\mathbf{k}) =\int d\mathbf{r}^3\,  e^{i \mathbf{k} \cdot\mathbf{r}} \mu_{s}(\mathbf r)$. {In this experiment, the system is a sphere of radius $R$ and mass $m$, for which we have} $\tilde \mu_{s}(\mathbf{k})={3m[{ \sin}(kR)-kR{ \cos}(kR)]}/{(kR)^{3}}.$
 By substituting $\tilde \mu_{s}(\mathbf{k})$, we obtain {a single diffusion constant independently from the direction of motion: }
\begin{equation}
\label{eta_CSL3}
\eta=\frac{6\rC^{4}\lambda m^{2}}{m_{0}^{2}R^{6}}\left [e^{-{R^2}/{\rC^2}}-1+\frac{R^{2}}{2\rC^{2}}(e^{-{R^2}/{\rC^2}}+1)\right ].
\end{equation}
The CSL-induced center of mass diffusion can be effectively described by a
{
stochastic force $f_\text{\tiny CSL}(t)$ with zero mean and correlation $\langle f_\text{\tiny CSL}(t)f_\text{\tiny CSL}(s)\rangle =S_\text{\tiny CSL}\, \delta (t-s)$, where $S_\text{\tiny CSL}= \hbar^{2} \eta$ is the CSL force spectral density. }

\begin{figure}[th!]
	\includegraphics[width=1\linewidth]{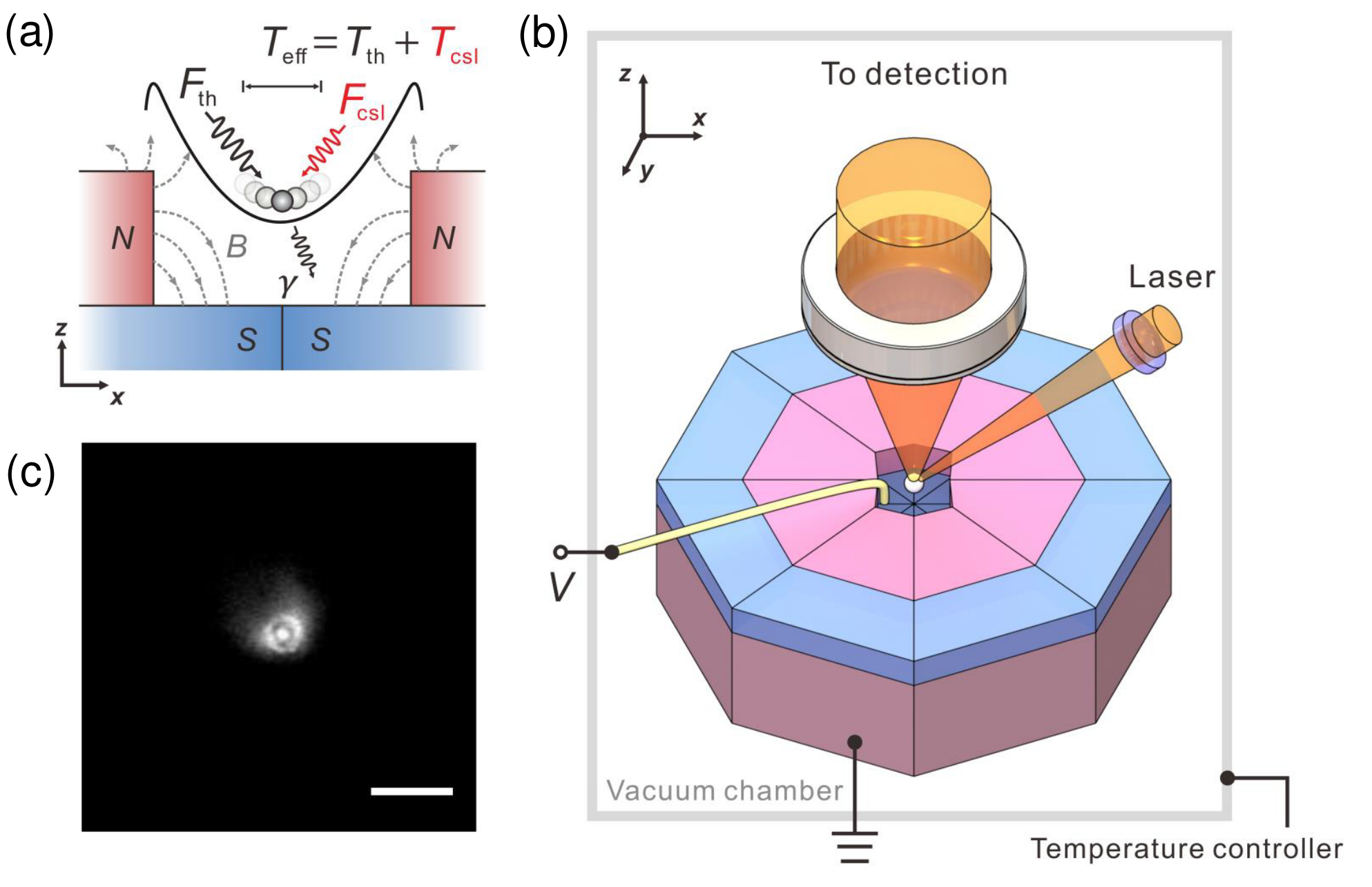}
	\caption{Basic concept of the experiment and setup. Panel (a): scheme of the oscillator dynamics. A micro-sphere trapped in a magneto-gravitational potential (black curve) is subject to Brownian motion with an effective
temperature $T_{\rm eff}$. In absence of the CSL collapse force $f_{\rm CSL}$, the thermal Brownian force $f_{\rm th}$ leads to an effective $T_{\rm eff}$ which is equal to the environment temperature $T_{\rm env}$. {When} $f_\text{\tiny CSL}$ {is} added, the effective temperature will rise by $T_\text{\tiny CSL}$. The smaller the oscillator damping rate $\gamma/2\pi$ the higher the effective temperature. Panel (b): experiment setup. A
diamagnetic
micro-sphere is levitated in a magneto-gravitational trap generated by a set of permanent magnets ({Red and blue} indicating the {N and S} poles, see Appendix \ref{Trap designing} for details). A laser is focused on the micro-sphere and
 an objective is used to collect the scattered light. An electrode is placed near the micro-sphere and an electric field is applied to determine the charge state of the micro-sphere. The whole setup is placed in a vacuum
chamber, whose temperature is controlled (see Appendix \ref{Trap designing} for details). Panel (c): optical image of the micro-sphere. The micro-sphere used in our experiment with a radius of $1\,\mu$m (scale bar being 5\,$\mu$m).
	}
	\label{setup}
\end{figure}

{We describe} the dynamics of {our mechanical system as a} {damped} harmonic oscillator subject to environmental and, potentially, CSL noises{. Dropping the label $i$, the motion in one direction of the center of mass of our system reads}
\begin{equation}
\label{EOM} m\ddot{x}+m \gamma \dot{x}  + k x = f_{\rm th}(t)+f_\text{\tiny CSL}(t),
\end{equation}
with $\gamma/2\pi $ the damping rate, {and} $k = m \omega_0^2$ the spring constant. The first term in the right hand side represents the thermal Brownian {force} noise, whose {correlations read} $\langle f_{\rm
th}(t)f_{\rm th}(s)\rangle = S_\text{th}\delta (t-s) $, where $S_{\rm th} = 2 \gamma m k_{\rm B}T_{\rm env}$ {the corresponding} power spectral density{, which is fully characterized by the environmental temperature $T_\text{env}$ \cite{rugar1995}}. For a system {in thermal} equilibrium, the {additional presence} of the
collapse force $f_\text{\tiny CSL}(t)$  leads to an increase of the temperature {of the system} \cite{Diosi2014}. {The effective temperature is thus }$T_{\rm eff} = T_{\rm env} + T_{\rm csl} ${, where} $T_{\rm csl}$, the CSL induced temperature {contribution,  satisfies} the relation $2 \gamma m k_{\rm
B} T_{\rm csl} =  \hbar^{2} \eta$.
Here{, one assumes that} $f_{\rm th}(t) $ and $f_{\rm csl}(t) $ {are independent}. Fig.~\ref{setup}~(a) shows an intuitive picture of the thermal equilibrium dynamics of the magneto-gravitationally levitated micro-sphere used in this experiment.

{It is clear that any other source of noise, such as for example that due to the  measurement back-action, also contributes to the total noise \cite{Clerk2010}. In Appendix~\ref{CSL dynamics} we discuss different possible noise processes involved in the experiment; however we take a conservative approach and consider all non-thermal noises as potential CSL noise in setting the upper bound.}

The {total}
power spectral density is defined as $S_{\rm total} = 2 \gamma m k_{\rm B}T_{\rm eff} $, which is calculated from the measured $T_{\rm eff}$. By subtracting  the thermal Brownian {contribution} $S_{\rm
th}$, we obtain the power spectral density of {all additional} force noises {$\delta S _{\rm total} =S _{\rm total} -S_{\rm th}$}. Therefore, $\delta S_{\rm total}
= 2
\gamma m k_{\rm B} \delta T$ provides the estimation of the upper bound of the CSL force as $\hbar^{2} \eta \leq  \delta S_{\rm total}$, with $\delta T = T_{\rm eff} - T_{\rm env} $ {denoting} the rise of the effective
temperature.
Note that, apart from a barely tunable material density $\sim m/R^3$ [cf.~Eq.~\eqref{eta_CSL3}], the ability to {test CSL is limited only by the accuracy in determining the thermal Brownian noise.} Different methods
for such noise signal sensing have been developed for {similar} purposes~\cite{Clerk2010,rugar2004,Kippenberg2012,huangpu1}.

\section{EXPERIMENT DESCRIPTION AND RESULTS}

The levitation {of diamagnetic systems} using magneto-gravitational forces has been {already performed with}  either superconductor \cite{Geim1999} or permanent magnets \cite{Slezak2018}. The magneto-gravitational trap used in {our} experiment {was} generated by a set of micro-machined NdFeB magnets with {octagonal bilayer} geometry {as} shown in Fig.~\ref{setup} (b). {In} Appendix \ref{Trap designing}, {we report details about the trap design.}  The
oscillator is a micro-sphere of polyethylene glycol, {whose} magnetic susceptibility is $-9.1\times 10^{-6}$ {and its density is $1.1 \times 10^3~$kg/m$^3$}. The micro-sphere is generated {using} a home-built nebulizer.  A 633\,nm laser is
focused on the droplet with a power less than 50$\,\mu$W, and the scattered light from the micro-sphere is collected with an objective. The position of the micro-sphere is {tracked} with a CCD camera, and its motion
{is recorded} in time domain with a photon detector. {To isolate the trap from external vibrations}, the trap is mounted on a {heavy} copper frame, which is suspended in a vacuum chamber by {means of} springs.
{Because the environmental temperature fluctuations contribute to the {measurement} uncertainty of {the} effective temperature of the oscillator, a double layer vacuum chamber and a PID temperature controller are used to maintain the environmental temperature stable.} {In this way, we achieved fluctuations {smaller than} 100\,mK with an environmental temperature set to 298\,K over the whole duration of the experiment ($\sim10^5-10^6$\,s).}

{We observed that for electrically charged particles, the dissipation in the experiment is higher than with no charge. This can lead to strong instability of the particles' motion, which makes them eventually escape from the trap. To avoid this, }
the charge on the
micro-sphere is eliminated by using ultra-violet light. {Subsequently,} the charge state is checked via a micro-electrode made of a gold wire of 40\,$\mu$m in diameter placed near the trapping centre. By applying a voltage $\sim50$\,V,
micro-spheres with radius less than 2\,$\mu$m can be easily pulled out of the trap if the net charge is nonzero (see Appendix \ref{experiments} for {further} details).
Even after having removed the charges, at room temperature {micro-spheres} with radius {smaller} than 500\,nm {are} found to escape the trap due to the thermal fluctuation {and the} limited  depth of the trap (see Appendix \ref{Trap
designing} for details).
The remaining particles were left in the trap in high vacuum
{
for several days. If the particle did not evaporate during this time, it eventually reached the equilibrium thermal distribution, which was observed to be stable within the measurement error. }

For micro-spheres {of} radius close to 1\,$\mu$m, direct optical image failed to provide a reliable estimation {of the size of the system}, and we determined {it through} the following two methods. In
the first method, we made use of the relation {between the} micro-sphere oscillation damping rate {due to the background gas collisions} and the pressure, {which reads} $ \gamma = (16/\pi)(P/ \nu R \rho) $ (holding for high pressures)\cite{optical_levitaion1}, with $P$ and $\nu$  the  pressure and the mean speed {of the background
gas} respectively. In this part of the experiment, the pressure was set to $P \sim 10^{-3}$\,mbar, so that the {damping} was fully {dominated} by the background gas; $\nu$ can be inferred from the environmental temperature, and by measuring $\gamma$ one obtains an estimate for $R$.

The second {method}  simply {relied on} the equipartition theorem. We
measured the oscillators displacement distribution, which follows {a} Gaussian distribution {$P(x)\propto \exp\left(-x^2/2\sigma^2\right)$, thus determining its} standard deviation $\sigma$. {The latter is related to the size of the particle through the} {energy} equipartition theorem:  $4 \pi \sigma^2 \mu
R^3 \omega_{0}^2 = 3 k_{\rm B} T_{\rm env}$, with {$\omega_0$ the resonant frequency of the oscillator}. The results from the two methods {are compatible}. The micro-sphere used in the  experiments, {whose results are described below}, has
radius $R = 1.0\,\mu$m, corresponding to $4.7\,$pg {and} with corresponding potential energy {which is} thousands of times large than the thermal energy $k_\text{B}T$.

\begin{figure}[ht!]
\centering
	\includegraphics[width=0.75\linewidth]{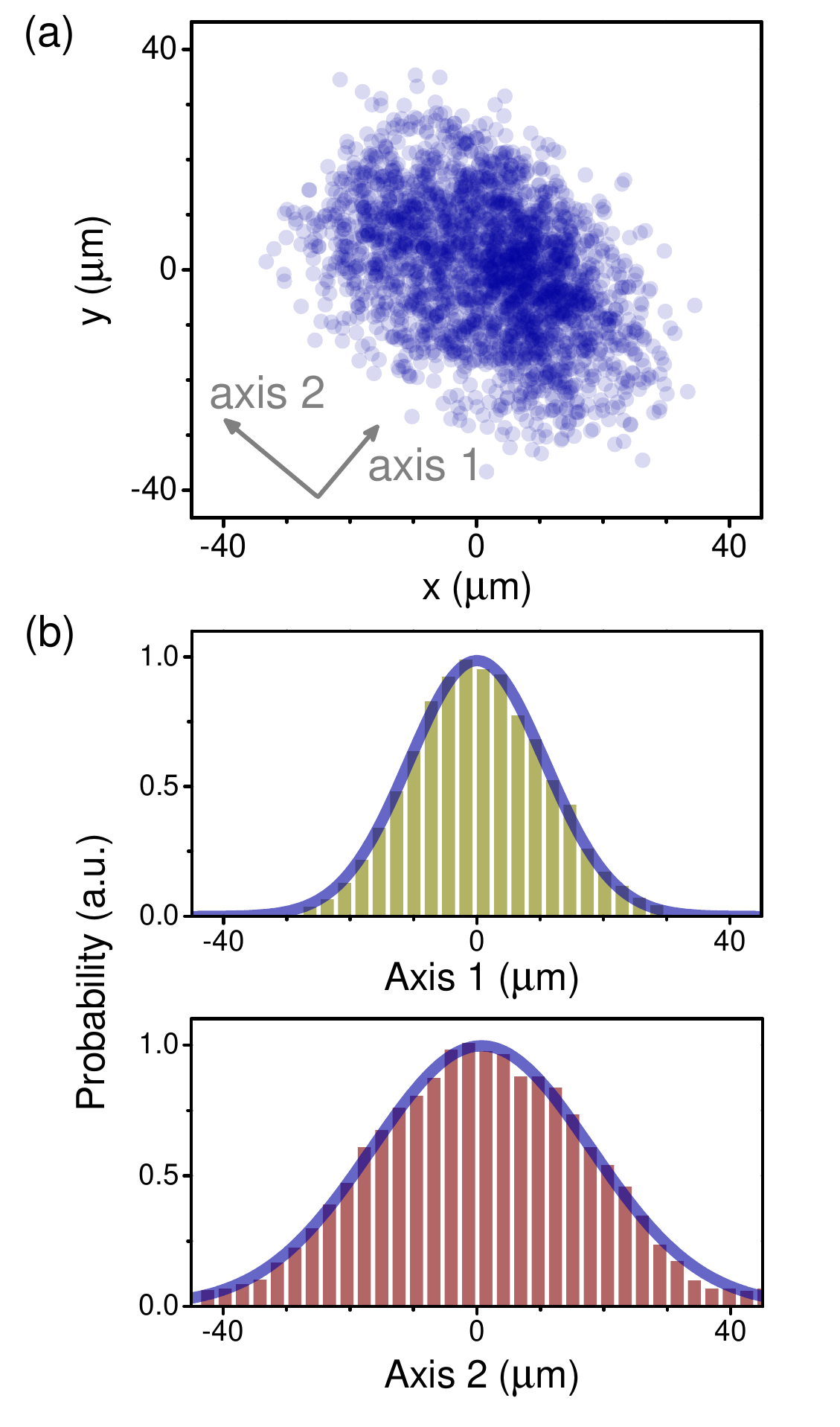}
	\caption{Measurements of the distribution in position of micro-sphere. Panel (a): typical position distribution in {the} $x$-$y$ plane. The data were measured under medium vacuum for few minutes. {The} axis 1 (short) and axis 2 (long) {of the distribution correspond} to the modes with resonance frequencies  $\omega_{1} = 12.9\,\rm Hz$ and $\omega_{2} = 9.3\,\rm Hz$, respectively.  Panels (b) and (c): displacement
distributions along axis 1 and axis 2, respectively. Data are fitted {with a} Gaussian distribution (blue curve), from which the effective temperature is determined.
	}
	\label{distribution}
\end{figure}

After successfully capturing the micro-sphere and eliminating {its} charge, we {proceeded to} the measurement of the effective temperature $T_{\rm eff}$ associated to the center of mass motion of the particle. {As a first step}, we set a medium vacuum ($P_{\rm MV} \sim
10^{-4}$\,mbar), and measured the position distribution of the micro-sphere in {the} $x$-$y$ plane. {A typical example of the}  measured data in a run of few minutes is plotted in Fig.~\ref{distribution}. {The distribution has} {an elliptical shape} {due to the asymmetry of the trap. The distribution}
can be fitted with {a} two-dimensional Gaussian distribution, whose long (axis 1) and short (axis 2) standard deviations are denoted by  $\sigma_1 $ and $\sigma_2$, respectively. The energy equipartition theorem{, which implies}
$\sigma_1/\sigma_2 = \omega_{1}/\omega_{2}$, is well satisfied within the measurement error, where $\omega_{1} = 12.9$\,Hz and $\omega_{2} = 9.3\,$Hz are the corresponding resonance frequencies (see Appendix \ref{nonlinearity} for {details on the displacement} power spectral density). The effective temperature is then calculated as $T_{\rm eff} = m \omega_{1}^{2} \sigma_1^2/k_\text{B}$ (equally, $ m \omega_{2}^{2} \sigma_2^2/k_\text{B}$). Since  at a
medium vacuum, the thermal Brownian noise from the background gas fully dominates the other noises, {we assume that} $S_{\rm total} =  S_{\rm th} $ and we use this relation {to calibrate} the environmental temperature as $T_{\rm env} = T_{\rm eff} $. Without loss of generality, we make use of mode 1 ($\omega_1/2\pi \approx  12.9$\,Hz) for the {subsequent} measurement.
\begin{figure}[t!]
	\includegraphics[width=0.9\linewidth]{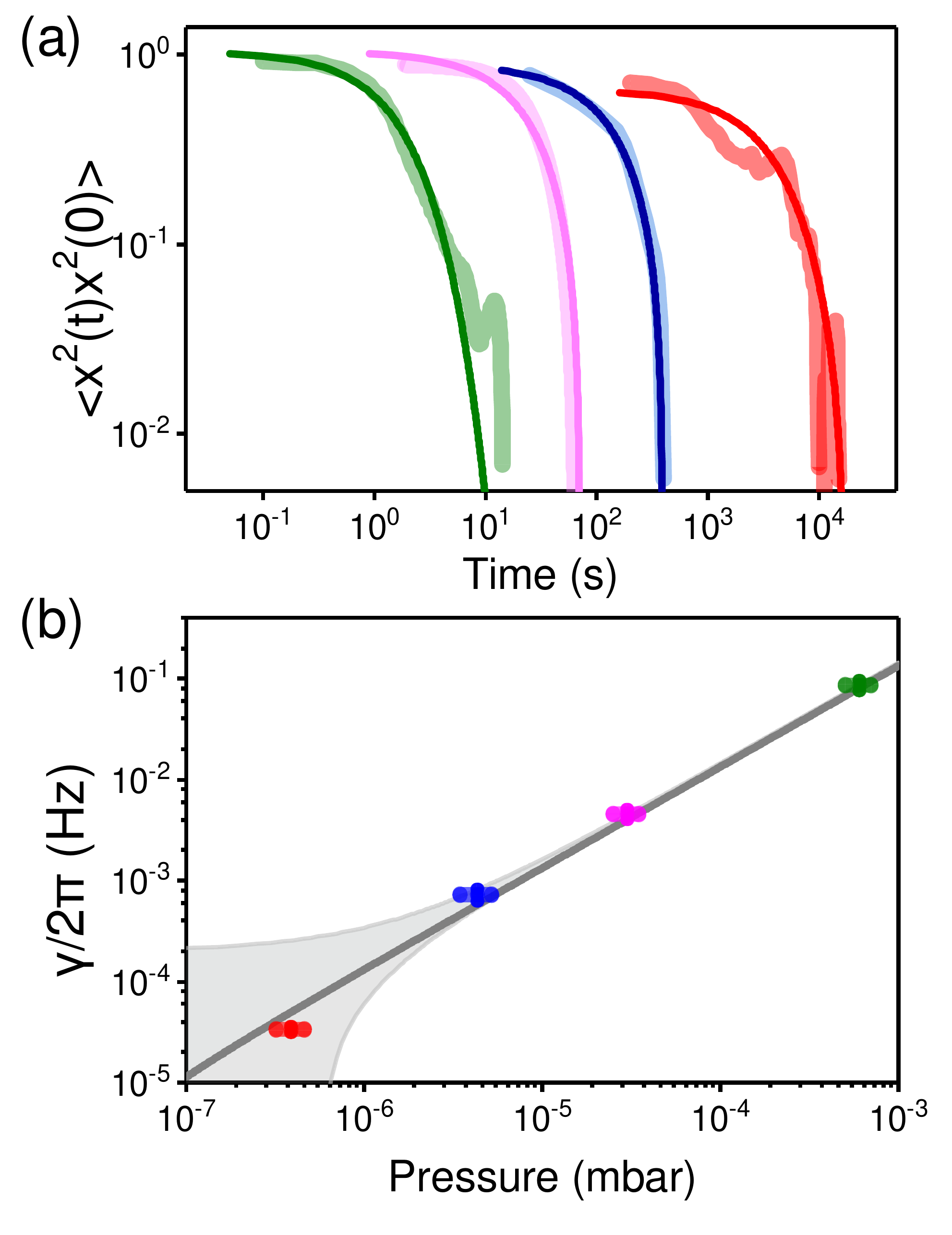}
	\caption{Measurement of {the} damping rate. Panel (a): autocorrelation functions of the oscillator energy measured at {the} pressures given in panel (b). {Data are fitted} with an exponential {decaying function} $\exp(-t/\tau)$. The decay time $\tau$ is determined
for each curve by the best fit. A total averaged time, typically $50$ times longer than $\tau $, was carried out for each curve to achieve good signal-to-noise ratio. Panel (b): dependence of the damping rate $\gamma/2\pi
=1/(2\pi\tau)$ on {the}
pressure, where $\tau$ is obtained from panel (a). The grey line is a linear fit, where the vertical and horizontal error bars are due to fitting error and the pressure imprecision of the vacuum {chamber}, respectively. The shaded grey area corresponds to the 95\% confidence band.
	}
	\label{dissipation}
\end{figure}

{Next, we determine the dissipation constant $\gamma$, }which is another key factor {for determining} the thermal Bronwnian force noise strength. {First}, we note that, for
$P < 10^{-5}$\,mbar, the {measured} power spectral density $S_{x}(\omega)$ shows {a} strong asymmetric character, {deviating} substantially from a {Lorentzian} shape, and is considerably broadened
compared to that estimated from the background gas. {Such a feature {is known} \cite{Dykman1992}  and is due to the nonlinearity of the trap.} {Therefore, to estimate $\gamma$} we {follow the prescription of \cite{rugar2001} and} make use of the
{energy} autocorrelation defined as $\langle X^2(t) X^2(0) \rangle$, with $X(t)$ the amplitude of {the} oscillation. {This method} is insensitive to the nonlinearity of the trap (see Appendix \ref{nonlinearity} for details). The measured autocorrelation curve is {then} fitted to the exponentially {decaying function} $\exp(-t/\tau)$, {from which we obtain the damping rate $\gamma = 1/\tau$.} Fig.~\ref{dissipation} (a) shows the measured {energy} autocorrelation {for different values of the pressure}. In particular, at the highest vacuum $P_{\rm HV} \approx 4 \times 10^{-7}\rm mbar
$, the measured decay time $\tau
\approx 4700$\,s {corresponds} to a damping rate $\gamma/2 \pi \approx   34 \,\rm \mu Hz$. We also find that the damping rate decreases linearly as the pressure decreases, which shows that the background gas
remains  the dominant {dissipative channel in the experiment}, as shown in Fig.~\ref{dissipation} (b). Combining the measured effective damping rate $\gamma/2\pi $ and temperature $T_{\rm eff}$, {we estimate a}  force sensitivity
of the oscillator in high vacuum {as} $\sqrt{S_{\rm total}} = 9.6 \rm \times  10^{-20}~N/\sqrt{Hz} $. {This value is comparable to that obtained from optical trapping \cite{optical_levitaion3}.}

\begin{table*}[ht!]
	\caption{\label{tab:tablel} {Upper bounds on the CSL collapse rate $\lambda$.} $\delta T$ is defined as the temperature increase $\delta T = T_{\rm eff} - T_{\rm env}$, with $T_{\rm eff}$ and $T_{\rm env}$ the effective temperature of the oscillator measured at high vacuum and the environment temperature, respectively.  $ \sqrt{\delta S_{\rm total}}$  is the measured additional force noise beyond the thermal force. {$\sigma_{\delta T}$ and} $\sigma_{\sqrt{\delta S_{\rm total}}}$ are the corresponding standard deviations at $95\%$ confidence level. {Finally, the upper bounds on }$\lambda$ at $95\%$ confidence level are the calculated for $\rC=10^{-7}\,$m and  $\rC=10^{-6}\,$m.  
 }
 \begin{ruledtabular}
\begin{tabular}{cc|cc|cc}

   \multicolumn{2}{c|}{ \ Excess temperature \ }  & \multicolumn{2}{c|}{Excess noise} & \multicolumn{2}{c}{Upper bound on the collapse rate}\\
\hline\hline
$ \delta T$&$\sigma_{\delta T}$&$ \sqrt{\delta S_{\rm total}}$ &$  \sigma_{\sqrt{\delta S_{\rm total}}}$&$\lambda$ ($\rC=10^{-7}\,$m)
  &$\lambda$ ($\rC=10^{-6}\,$m)\\
    $6.5\,K $     & 40\,K    &$1.3\rm \times  10^{-20}N/\sqrt{Hz} $    &   $3.3 \rm \times  10^{-20}N/\sqrt{Hz} $  & $10^{-6.4}{\,\rm s} ^{-1}$ & $10^{-7.4}\,{\rm s} ^{-1}$
\end{tabular}
\end{ruledtabular}
\end{table*}

By comparing the power spectral densities at medium and high vacuum we find the {upper bound on the collapse} rate $\lambda$;  the main results are summarized in Table~\ref{tab:tablel}. For medium vacuum, the background gas is coupled to the system, thus maintaining the temperature of the system at  equilibrium with the environmental one. On the contrary, in high vacuum condition, the gas decouples, and thus any potential CSL contribution is not dissipated, thus imposing an effective temperature which is higher than $T_\text{env}$. To bound the CSL effect we proceed as follows. The {power spectral density of non-thermal forces is estimated via}  $\delta
S_{\rm total} = 2 m \gamma k_{\rm B}\delta T_{} $,
where $\gamma$ is  measured at high vacuum, $\delta T =
T_{\rm eff} - T_{\rm env} $, with $T_{\rm eff}$ being calculated from the standard deviation of {the} position distribution $\sigma$ at high vacuum and {with} $T_{\rm env}$ measured at medium vacuum. We obtain the upper bound {$\sqrt{\delta S_{\rm total}}<  3.3 \times 10^{-20}\rm N/\sqrt{Hz}$}
 at the 95\% confidence level (see Appendix \ref{error} for
details about measurements of $T_{\rm eff}$ and error estimation). {Accordingly, the bound on} $\lambda$ is calculated {through Eq.~}\eqref{eta_CSL3}.

\begin{figure}[th!]
	\includegraphics[width=1\linewidth]{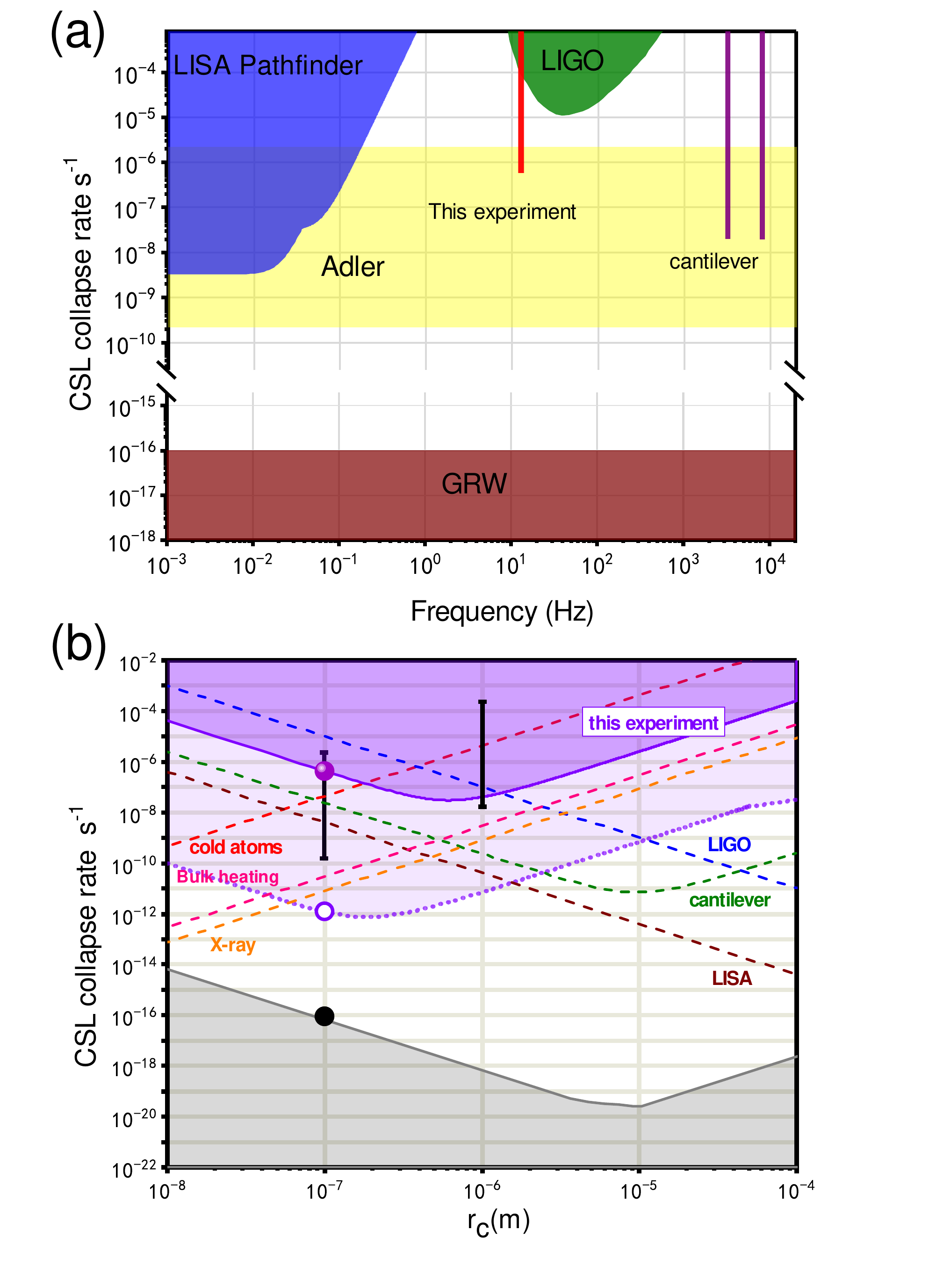}
	\caption{{Exclusion plot of the CSL parameters.}  Panel (a): upper bounds on the CSL collapse rate $\lambda$ from mechanical-based experiments, {for} $\rC = 10^{-7}\,\rm m$. Red line: upper bound given by our experiment at the $95\%$ confidence level. Blue and green regions, and {purple}
lines: exclusion regions obtained from LISA Pathfinder~\cite{LISA1,LISA2,Rotational1,gravitation results1,gravitation results2}, Advanced LIGO~\cite{LIGO1,gravitation results1} and millikelvin cantilever {experiments}~\cite{Vinante2016,Vinante2017}, respectively.  The
yellow region is the {range proposed by Adler for $\lambda$} \cite{Adler2007}.  The claret region is the value proposed by Ghirardi, Rimini and Weber (GRW)~\cite{GRW} {and it works as a lower theoretical bound}.  Panel (b): upper bounds in the $\lambda-\rC$ plane given by our experiment, compared with the best experimental
upper bounds reported so far as well as proposed theoretical lower bounds. The purple solid line {and corresponding shaded region}: upper bound and exclusion region given by our experiment. Purple dotted line: upper bound estimated with parameters $R=0.3\,\mu$m, $\gamma/2\pi$=10$^{-6}$\,Hz and $\delta T=10$\,mK. At $\rC=10^{-7}$\,m, values of the collapse rate $\lambda$ {obtained by this experiment} and its future possible improvement are marked by a purple solid dot ({$\lambda=10^{-6.4}\,$s$^{-1}$}) and a purple hollow dot ({$\lambda=10^{-11.9}s^{-1}$}) {respectively}. The blue, green, claret, red, pink and orange dash lines represent the upper bounds given by LIGO, cantilever, LISA, cold atoms \cite{ultra-cold atom2}, bulk heating \cite{Adler2018} and X-ray emission \cite{X-ray,X-ray2}, respectively. Dark bars: the theoretical values suggested by Adler. Black dot and grey region: the GRW value and the theoretical lower bound \cite{Interf1,Interf2}.
	}
	\label{lambda}
\end{figure}

Fig.~\ref{lambda}(a) {compares the excluded values of $\lambda$ at different frequencies for $\rC=10^{-7}\,$m. In particular, our experiment improves by more than two orders of magnitudes the bound posed by Advanced LIGO~\cite{LIGO1,gravitation results1} at the same frequency.}

The upper bound
provided by this experiment also partially excludes the range of values of collapse rate suggested by Adler for $\rC = 10^{-7}\,$m {\cite{Adler2007}}, and almost {entirely} excludes it for $\rC= 10^{-6}\,$m. We also estimated the performances of this {experiment by} using parameters that are more favourable for CSL testing {and which are potentially achievable with our experiment by working at cryogenic condition and for 10$^{-8}$\,mbar pressure:} $R=0.3\,\mu$m,  $\gamma/2\pi=10^{-6}\,$Hz and $\delta T=10$\,mK. A negative result would imply $\lambda \leq 10^{-11.9}\, \rm s^{-1}$ for $\rC=10^{-7}\,$m, which would
fully rule out Adler's suggestion. {The comparison of our experimental upper bound and the hypothetical upper bound with the} strongest bounds reported in the literature are shown in Fig.~\ref{lambda}(b), together with the theoretical values for the collapse parameters.


\section{ SUMMARY AND DISCUSSION }
{Levitated oscillators have been recently proposed as suitable systems for the collapse model testing }\cite{levitation1,levitation2,levitation3,levitation4,gravitation results3}. {Here,} we
demonstrated {that an experiment based on a magneto-gravitational levitated miro-oscillator can place important bounds on the collapse parameters although operating at room temperature}. {We obtained a new} upper bound, which is a significant improvement over previous {results in}
the same frequency range and {it partially probes} Adler's {theoretical proposal}. The system reported here {shows a} great potential, {which would be fully expressed at cryogenic temperatures, where an improvement of several orders of magnitude in bounding the collapse noise is expected.}

The performance of {the} current experiment {at room temperature} is mainly limited by three factors, which {eventually} could be improved significantly at {lower} temperatures. {First}, the effective temperature measurement precision is worse than tens Kelvins but is expected to
reach mK under cryogenic conditions. {Differently} from other kinds of levitated micro-oscillators, such as electrical \cite{electric_levitation}, optical  \cite{optical_levitaion1,optical_levitaion2,optical_levitaion3} and magnetical levitation \cite{Magnet205,Magnet206,Pino}, our magneto-gravitational trap is fully passive {with no energy inputs}. {Thus, it} is naturally suitable for low temperature condition. (In principle laser generates an addition force noise. However, the laser intensity is weak at room temperature. Its impact at cryogenic temperature is still to be evaluated.) {Second, } the
minimum radius of the micro-sphere in this experiment is currently limited by {the} thermal energy, {thus,} at low temperature, much smaller micro-sphere could be stably trapped and {lead to higher precisions in detection}. {The third potential of improvement is dissipation, which} is observed to be constrained by the pressure. Room-temperature experiments show that higher vacuum does not lead to significantly improvement in
dissipation \cite{Slezak2018}, {since eventually other dissipative channels will contribute at lower pressures.} However, it is yet to be {explored} whether dissipation can dicrease
 at a much lower temperature environment. This work opens a new door for the precise study of collapse models and may provide
promising avenues towards breakthrough discoveries in the future.

\begin{acknowledgments}
We thank L. Di\'{o}si for helpful comments. We thank Xuan He, Zhangqi Yin, Zhujing Xu, Jingwei Zhou, Fei Xue and Xing Rong for helpful discussion. This work was supported by the National Key R\&D Program of China (Grant No. 2018YFA0306600), the National Natural
Science Foundation of China (Grant No.~61635012, No.~11675163, No.~11890702, No.~81788101, No.~11761131011, and No.~11722544),  the CAS (Grant No. QYZDY-SSW-SLH004, Grant No. GJJSTD20170001), the Fundamental Research Funds
for the Central Universities (Grant No. 021314380149), and the Anhui Initiative in Quantum Information Technologies (Grant No. AHY050000). {MC and AB acknowledge financial support from the H2020 FET Project TEQ (Grant No. 766900). AB acknowledges financial support from the COST Action QTSpace (CA15220), INFN, and the University of Trieste.}

\end{acknowledgments}

\textbf{Note Added}: After this work was completed, we became aware of similar independent work by Pontin \textit{et al.} \cite{Pontin2019}.

\appendix

\section{Calculation of fluctuation dynamics} \label{CSL dynamics}
The system was {modeled} by a classical mechanical oscillator with the motion described by the Langevin equations{, which, }{in vectorial form}{, read}:
\begin{equation}\label{EOM1}
m  \ddot{\mathbf{x}}+m \mathbf{\Gamma} \dot{\mathbf{x}}  + \mathbf{K}\mathbf{x} + \textit{o}(\mathbf{x}^3) = \mathbf{f}_{\rm th}(t)+\mathbf{f}_{\rm CSL}(t)+\mathbf{f}_{\rm add}(t),
\end{equation}
where
{$\mathbf{x} = (x,y,z)$, $m$ is mass of the oscillator,  $\mathbf{\Gamma} $   is the {damping rate} diagonal matrix {with elements}  $\gamma_{ii}$ {($i$ corresponding to x, y and z)}. {When} the
background gas damping dominates, $\gamma_{ii}$ are isotropic:  $\gamma_{ii}=\gamma$. Similarly, $\mathbf{K}$ is the diagonal matrix of the effective spring constants with element $k_{i} = m
\omega_{i}^2$  and $\omega_{i}$ is the resonance frequency of the oscillator {along the $i$-th axis}.} {$\textit{o}(\mathbf{x}^3)$ includes the higher order terms beyond the linear oscillator, such as Duffing nonlinearity $\alpha_i x_i^3$ and
nonlinear couplings between different motions as $\beta_{i,j} x_i x_j^2$ etc\cite{Nonlinear2009}.} The right side of the equation is a sum of force noises. {They include} the thermal fluctuations ${f}_{\rm th}(t)$, possibly the CSL
 induced stochastic force ${f}_{\rm CSL}(t)$, and {all the additional contributions, e.g.~those due to the optical measurements, mechanical vibrations ecc.}

Considering the motion in a single direction and dropping the direction label $i$, we estimate the three contributions to the noise in the system.  {The first one,} the thermal force noise, was estimated by using {the} fluctuation dissipation theorem which gives
 the relation $\langle f_{\rm th}(t) f_{\rm th}(0)\rangle = 2 m \gamma k_B T_{\rm env}  \delta (t)$, where $T_{\rm env}$ is environmental temperature. Equivalently, {its} strength can be described by the power spectral
density $S_{th}(\omega)= 2 m \gamma k_B T_{\rm env} $. {The second contribution, $f_\text{\tiny CSL}$ has been described in the main text.} Within the third contribution, $f_\text{\tiny add}$,  the optical force noise is the dominant one. It can be written as $f_{\rm opt}(t)=  f_{\rm int}(t)+f_{\rm sc}(t)$. The first term $f_{\rm int}(t)$ is the classical optical force due
to intensity fluctuations, including both {those from the} intensity fluctuation $\delta I(t)$ and the position fluctuation of the light {position} $\textbf{x}_{\rm opt}(t)$ relative to the {center of the} magneto-gravitational trap.  The
illumination light intensity fluctuation induced force can be expressed as $-\alpha'   \nabla \xi [\textbf{x}_0]  \delta I(t) /4$  and the light spot position fluctuation induced force  as $- \alpha' I \nabla (\nabla \xi
(\textbf{x}_0)\cdot \delta \textbf{x}_{\rm opt}(t))$, with $\textbf{x}_0$ the trapping position, $\xi (\textbf{x}_0)$ the normalized light field distribution function, and  $I$ the average intensity of the illumination
light. The second term $f_{\rm sc}(t)$ is the stochastic force due to photon scattering. {An additional contribution to $f_\text{\tiny add}$ is} the parametric noise that is generated from the illumination light intensity, which {leads} to a fluctuation of the spring
constant $k_{}$ via optical force, and is proportional to $ \delta I(t)$~\cite{parametric cool}.

We solved the Fokker-Planck equation for the probability density to obtain the {statistical} behavior of the system. To  this {end}, the Langevin equations of motion in {a} single direction {are} written as:
\begin{equation}
m\ddot{x}+ m  \mathbf{\gamma}\dot{x} + m\omega^{2}_0x + m\omega^{2}_0\zeta (t) x = f_{\rm total}(t),
\end{equation}
where the parametric fluctuation $\zeta(t)$ was approximately taken as {a} white noise {satisfying} $\langle \zeta(t) \zeta(0)\rangle = \varsigma\delta (t)$. $f_{\rm total}(t)=f_{\rm
th}(t)+f_\text{\tiny CSL}(t)+f_{\rm opt}(t)$ is the total force noise and it was also {assumed to be white}: $\langle f_{\rm total}(t) f_{\rm total}(0)\rangle = 2 m \gamma k_B T_{\rm eff}  \delta (t)$, with $T_{\rm eff}$ the effective temperature. It is
noted that  $f_{\rm total}(t)$ and $\zeta(t)$ are not strictly independent, because both  contain the contribution from the illumination light intensity fluctuation $\delta I (t)$. However, such a contribution in
$f_{\rm total}(t)$ is small, and so we took  the total force noise and the parametric noise as {approximately} independent. {Given this}, we write the Langevin equations of motion as {follows} \cite{Kampen1976}:
\begin{eqnarray}
\begin{aligned}
&  \D x=\frac{p}{m}\D t,  \\
&  \D p=(-m\omega_{0}^{2}x-\gamma p)\D t+  \\
& \sqrt{2m\gamma k_{B}T_{\rm eff}}\D X+m\omega^{2}_0 \varsigma x\D Y,
\end{aligned}
\end{eqnarray}
with $\D X$ and  $\D Y$ two independent random variables {with} Gaussian distribution. Setting the energy of the oscillator as $\varepsilon = p^2/2m + k x^2/2$, with high quality factor $Q=\omega_0/\gamma$, the  Langevin
equations of motion leads to the Fokker-Planck equation {for the} probability density $P(\varepsilon,t)$ reads

\begin{equation}\label{Fokkereq}
\begin{aligned}
\frac{\partial P(\varepsilon,t)}{\partial t}=&-\frac{\partial}{\partial \varepsilon}[(-\gamma \varepsilon+\gamma k_{B}T_{\rm eff})P(\varepsilon,t)]\\
&+\frac{\partial^{2}}{\partial \varepsilon^{2}}[(\gamma k_{B}T_{\rm eff}\varepsilon+\frac{1}{4}\omega_{0}^{2}\varsigma^{2}\varepsilon^{2})P(\varepsilon,t)].
\end{aligned}
\end{equation}
For a stationary probability distribution $\partial P(\varepsilon,t)/\partial t = 0 $, and Eq.~\eqref{Fokkereq} is:
\begin{equation}
P(x^2) = \frac{\gamma+\frac{1}{2}\omega_{0}^{2}\varsigma^{2}}{\gamma k_{B}T_{\rm eff}}\left (1+\frac{m\omega_{0}^{4}\varsigma^{2}}{2\gamma k_{B}T_{\rm
eff}}x^{2}\right)^{-\frac{2(\gamma+\omega_{0}^{2}\varsigma^{2})}{\omega_{0}^{2}\varsigma^{2}}}.
\label{prob_disA10}
\end{equation}
This distribution was measured experimentally. It is noted that, for the limiting case  $\varsigma \rightarrow 0$, the expression {in} Eq.~\eqref{prob_disA10} approaches the Gaussian distribution.

\section{Design of the magneto-gravitational trap} \label{Trap designing}
\begin{figure}[h!]
\includegraphics[width=1\linewidth]{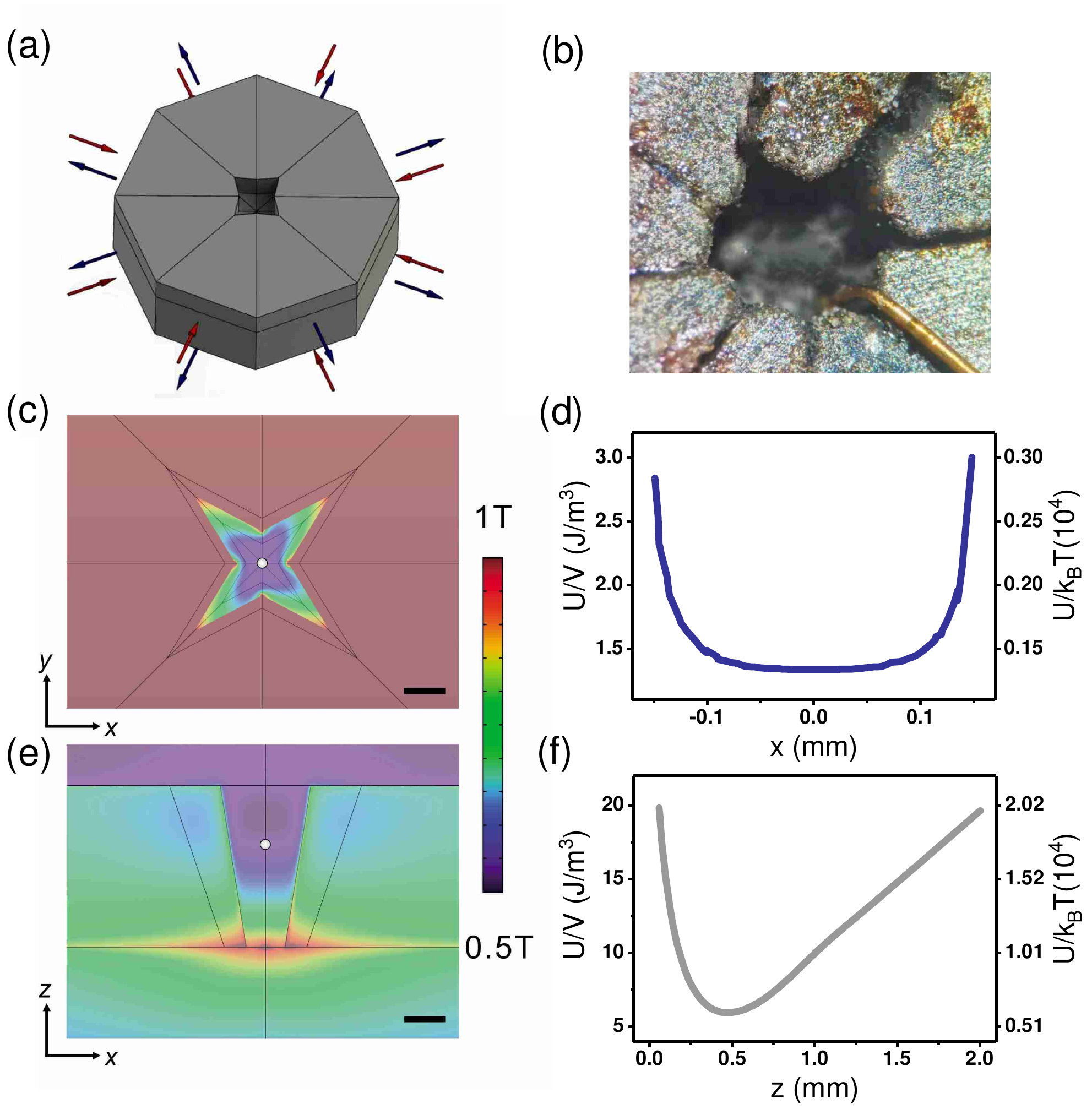}
\caption{Design of the magneto-gravitational trap. Panel (a): schematic diagram of the trap. It consists of two layers of magnets with opposite magnetization; the arrows point in the direction of the nord magnetic pole. Panel (b): image of the central portion of the trap used in this experiment with an area of about $1 \rm mm \times 1 \rm mm$.
Panels (c) and (e): geometry and simulated magnetic field strength $|\bm{B}|$ of the trap in $x$-$y$ and $z$-$x$ planes (scale bar equals 200\,$\mu$m). The grey point shows
the position where the micro-sphere is trapped. The micro-sphere is made of polyethylene glycol material with density $\mu = 1.1\times
10^3~\rm kg/m^3$  and a magnetic susceptibility $\chi = -9.1 \times 10^{-6} $.
Panels (d) and (f): calculated magneto-gravitational potential {energy density (left axes)} as function of $x$ and $z$ coordinates respectively. {On the right axes, we report the corresponding energy (divided by $\mathrm{k_{B}T_{env}}$) {for a micro-sphere of diameter 2$\mu$m}}.
}
	\label{trap}
\end{figure}
The potential energy density of a small diamagnetic micro-sphere in a magneto-gravitational trap under {an} illumination light field can be written as \cite{Berry1997}:
\begin{equation}\label{Potential}
U(\mathbf{x})=-\frac{\chi }{2 \mu_0} V \left |\bm{B}(\mathbf{x})\right |^2  + mgz + \frac{\alpha' }{4 } I V \xi (\mathbf{x}) + U_0.
\end{equation}
Here, the first term is the diamagnetic potential, with $\chi$ and $v$  the magnetic susceptibility and volume of the micro-sphere; the second term is the gravitational potential, with $m$ the mass of the micro-sphere
and $z$ {is} taken {opposite} to the direction of gravity; the third term is the optical gradient force, with $\alpha'$ the real component of the polarisability, $I$ the light field intensity proportional to the light power, and $\xi
(x)$ the normalized light field distribution function. The conditions that a diamagnetic micro-sphere can be stably trapped in {the} equilibrium position $\mathbf{x_0}$ are:
\begin{eqnarray}
\label{stable} && {\rm \textbf{F}}(\bm{x_0}) = -\nabla \mathit{U} (\mathbf{x_0}) = 0, \\
&& \nabla \cdot{\rm \textbf{F}}(\mathbf{x_0}) < 0,
\end{eqnarray}
{with} $ \textbf{F}(\mathbf{x})$  the total force of the potential. Near the equilibrium position $\mathbf{x_0}$, the potential can be {approximately} expressed in quadratic form {with respect to} the displacement $\mathbf{x}$ from
$\mathbf{x_0}$ as $U(\mathbf{x} +\mathbf{x_0})\approx \Sigma _{i,j}  \frac{\partial^2 U(\mathbf{x_0})}{\partial x_i\partial x_j}  x_i x_j$ ($i,j = x,y,z$), which can be put into a diagonalised form {as} the sum of three
independent harmonic oscillators,
\begin{equation}
\label{quadratic form} U(\mathbf{x}+\mathbf{x_0})\approx \sum_{i} \frac{1}{2} k_i x_i^2,
\end{equation}
where $k_i$ {with $i=x,y,z$,} are the effective spring constants, leading to the characteristic frequencies of the oscillators  $\omega_i = \sqrt{k_i/m}$. The constant term $U_0$ is dropped for convenience. The optical
field will also generate an effective potential via the optical force, however such an {effect} is much smaller than $U(\mathbf{x})$ and can be {neglected}. Hence, in the trap design, only the magnetic and the gravitational energies were taken
into account. The potential function was calculated using a finite element simulation and the result is plotted in Fig.~\ref{trap}.

\begin{figure}[h!]
	\includegraphics[width=1\linewidth]{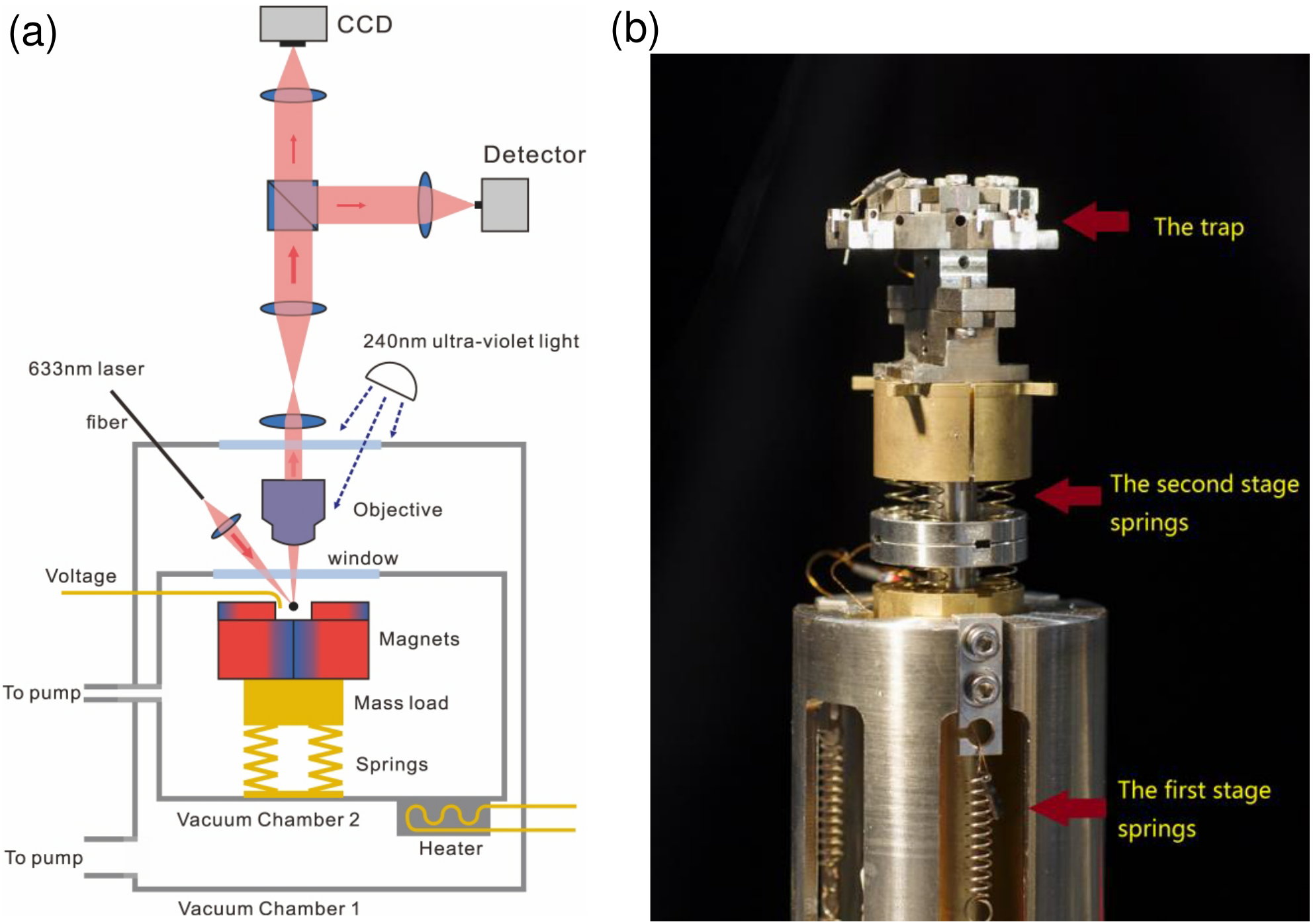}
	\caption{Experimental setup. Panel (a): a double layered vacuum chamber is used for temperature control. The environmental temperature of the inner chamber is maintained constant by PID feedback, and is kept slightly higher than the room temperature
by using a heater in the outer chamber. Ultra-violet light is used to eliminate the charge on the micro-sphere and an electrode in the trap in used to test the charge by applying a voltage. The 633\,nm laser is applied to the micro-sphere, and
the scattered light is collected by a CCD camera; the position and motion of the micro-sphere is recoded by a photon detector. The set of magnets forming the trap is mounted on a heavy copper frame, which is suspended via springs to
isolate external vibrations. Panel (b): image of copper frame with the vibration isolation system consisting of a two-stage spring-mass based suspension.}
	\label{experiment detail}
\end{figure}

\section{Experimental setup and micro-sphere generation}
 \label{experiments}

The {experimental} apparatus is shown in Fig.~\ref{experiment detail}(a): the magneto-gravitational trap is held {in a vacuum system} by specially designed {springs}, with temperature of the inner chamber monitored and controlled
to be slightly above the room temperature, and the pressure controlled by a turbomolecular pump of tunable rotation speed. A CCD camera was used to detect the position of the micro-sphere, {the} magnification
$\mathcal{M}$ of the detection optics being calibrated by a standard micro-structure so that the displacement in the $x$-$y$ plane of the micro-sphere {is} $\mathbf{x} = \mathbf{x^{\prime}} /\mathcal{M}$, where $\mathbf{x^{\prime}}$
is the displacement of the micro-sphere image read {out} by the CCD camera. In this way, the thermal distribution was obtained.

A photodetector was used to detect the position-dependent scattering light intensity $I_{\rm sc}$, which is proportional to the illumination light $I$ as $I_{\rm sc}\propto I \xi (\textbf{x}$+\textbf{x$_0$}). Since the
thermal motion is much bigger {than} the wavelength, such  detection scheme is efficient. The power spectral density {in the position} $S_{x}(\omega)$ is then calculated from the output photon detector voltage: $S_{x}(\omega) \propto
S_V(\omega)$, with $S_V(\omega)$ the power spectral density of the output voltage. For high quality
factor oscillators, the detection nonlinearity does not influence the results.

In order to eliminate the influence of the external vibration, the whole experimental setup is {first} mounted on an optical table with air legs, and a two-stage spring-mass based suspension is used to further isolate the vibrations, as shown in Fig.~\ref{experiment detail}(b). The resonance frequency of the first stage (the second stage) in $x$-$y$ plane is about 1.5\,Hz (4\,Hz), and the mass of the first stage is designed to be much heavier than the mass of the second one. We used a very thin wire with diameter about 40\,$\mu$m to {apply} the electric field which was used to pull the micro-sphere, and the wire was mechanically bounded on the first and then second stage before going to the trap, so that vibrations transmitted through the wire to the trap were effectively suppressed.

The micro-sphere used in our experiment is a small polyethylene glycol 400 droplet. To generate such a droplet with desirable diameter, we first mixed polyethylene glycol 400 with dibutyl sebacate (DBS) and ethanol in the
proportion of $1:27:1000$ (volume ratio). Subsequently, droplets of the suspension were sprayed into the trap  using a home built piezo-atomizer at atmospheric pressure. Ethanol rapidly evaporated {after some} seconds and a droplet with
typical diameter of 3 to 7\,$\mu \rm m$ {was} obtained. Next, a moderated voltage of about a few tens of Volts was applied while the displacement of the droplet was monitored, and an ionizing radiation source (Americium-241) was brought
near the droplet. After exposing the droplet to the radiation for a few seconds, the charge on it changed randomly. Once a positively charged droplet was obtained, the pressure was gradually decreased to $10^{\rm -6} \,\rm mbar$
for one day, and then DBS fully evaporated and the diameter of the micro-sphere {did} not change anymore. Next, a ultra-violet light was used to slowly eliminate the positive charge until the droplet became fully neutralized.
This was determined as follows: for a micro-sphere with only a few electron charges, jumps in voltage-displacement response became clear, and eventually the responses dropped to zero when the net charge went to zero by applying a voltage larger than  $50$\,V.  We also observed that the charge state was stable in vacuum ($P<10^{\rm -4} \,\rm mbar$) for {a} very long time (tens of days or even longer).

\begin{figure}[h!]
\includegraphics[width=0.9\linewidth]{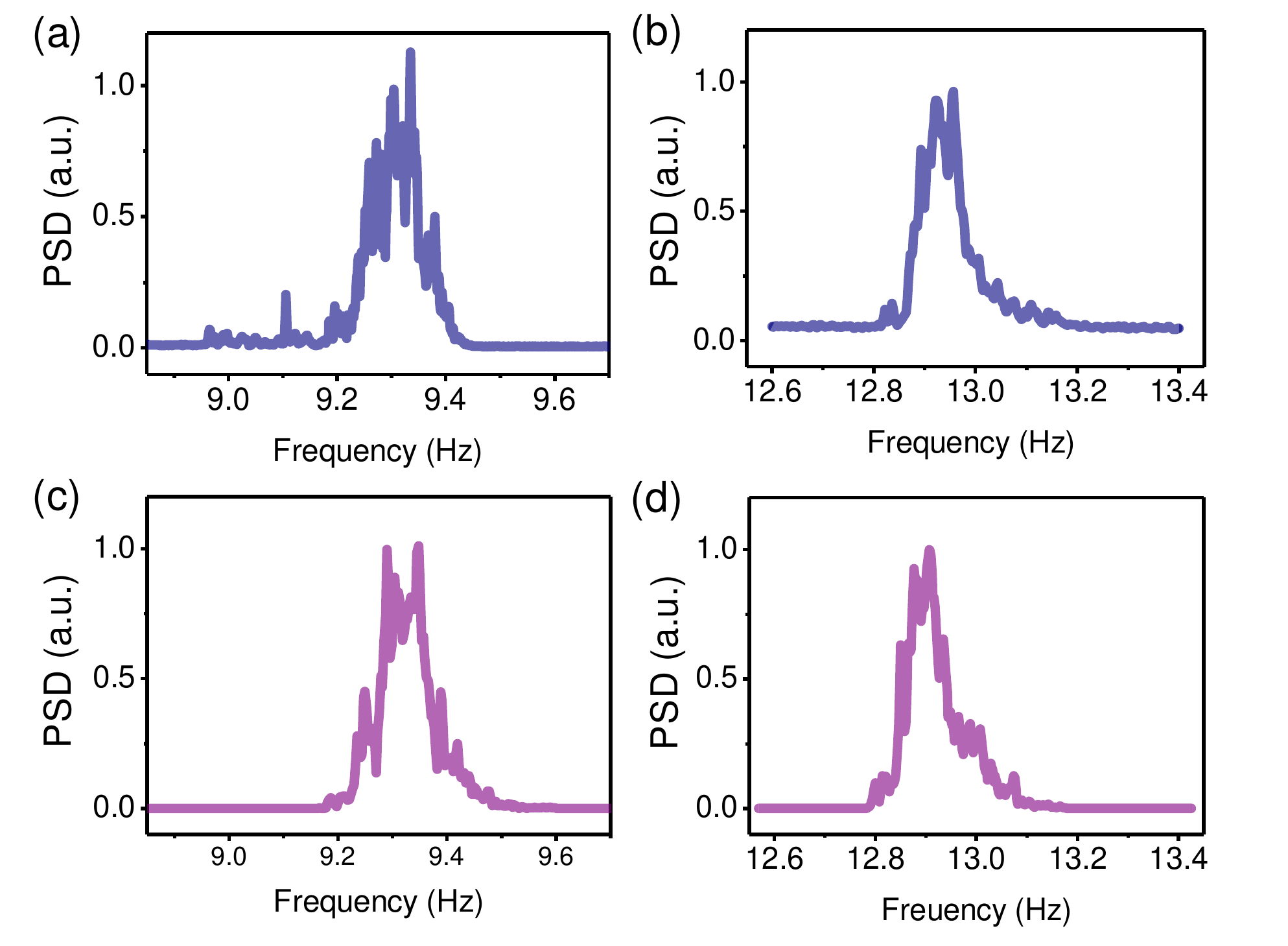}
\caption{Power spectral density (PSD) of the displacement $S_{\rm x}(\omega)$ under high vacuum. Panels (a) and (b): measured data of two different oscillation modes corresponding to the resonance frequencies $\omega_2/2\pi
\approx
9.3\,\rm Hz$ and $\omega_1/2\pi \approx  12.9\, \rm Hz$; the  full width at half maximum of the peak turns out to be much larger than $\gamma/2\pi $ and asymmetric, which can be explained by the nonlinearity of the trap. Panel (c) and
(d): numerical
simulations of panels (a) and (b) by introducing a nonlinearity, where the nonlinear coefficients are adjusted so that simulation and experiment agree with each other.}
	\label{PSD}
\end{figure}

\section{Influence of nonlinearity on measurement results} \label{nonlinearity}
The nonlinear term in Eq. \eqref{EOM1} becomes important for a motion with large amplitude. For simplicity, we consider the
term $\alpha x^3$ but temporarily we omit the coupling terms $\beta _{i,j} x_{i}^2 x_{j}$; the oscillator {then becomes} a Duffing oscillator with the
following equation of motion \cite{Nonlinear2009}:
\begin{equation}
\label{EOM2} m\ddot{x} +m \gamma \dot{x}  + k x + \alpha x^3 = f_{\rm total}(t),
\end{equation}
where $\alpha$ is the Duffing constant. One of the important effects of nonlinearity is a frequency shift and broadening that are proportional to the thermal fluctuation $\alpha k_{B}T_{\rm eff}$ \cite{nonlinear1980}. When such
a nonlinear thermal {broadening} becomes larger than the damping rate $\gamma$, the power spectral density shows a non-Lorentz character \cite{Dykman1992}. Hence, in the thermal nonlinear regime, the damping rate $\gamma/2\pi$
cannot be obtained by measuring the full width at half maximum based on the power spectral density, which is commonly used {with an} harmonic oscillator. Instead, we notice that the change of energy over time is still the same as
that of a harmonic oscillator. This is because the reduction of energy in the damping process results from the dissipation via {the} kinetic energy $p^2/2m$, while {the} nonlinearity only modifies the potential energy {and}
preserves energy conservation \cite{rugar2001}. {Therefore,} we extract $\gamma$ from the energy autocorrelation as described below.

From Eq.~\eqref{EOM2}, we first write  equations of motion for position and momentum, without the fluctuation $f_{\rm total}(t)$, i.e.
\begin{equation}\begin{aligned}
\D x&=\frac{p}{m}\D t,\\
\D p&=(-k x-\alpha x^{3}-\gamma p)\D t.
\end{aligned}\end{equation}
Then the change of the total energy of the oscillator follows
\begin{equation}
\label{energy decay} \D \varepsilon=\frac{\partial \varepsilon}{\partial x}\D x+\frac{\partial \varepsilon}{\partial p}\D p=-\frac{\gamma p^{2}}{m}\D t.
\end{equation}
Next, we consider a short period during which the dissipation is negligible so the motion of the system can be written as
\begin{equation}
x(t)=X(t)\left [ \cos(\omega t)+\frac{\kappa X^{2}(t)}{12}  \cos(3\omega_{} t)+o(\kappa^2 X^{4}(t))\right ].
\end{equation}
Here, $X(t)$ is a vibrational amplitude that is slowly varying, $\kappa = 3 \alpha / 8 m \omega_0^2$ and $\omega$ is an amplitude dependent oscillation frequency which shifts from {the} resonance frequency $\omega_0$ as:
\begin{equation}
\omega =\omega_{0}(1+\kappa X^{2}(t)).
\end{equation}
 As $X(t)$ goes to zero, we have $x(t)\approx X(t) \rm {cos}(\omega t)$, as expected.

Then, we define the average kinetic energy $E_K$ and average potential energy $V$  as
\begin{equation}
E_K=\frac{1}{\tau}\int_{0}^{\tau}\frac{1}{2}m\dot{x}^{2}(t)dt,
\end{equation}

\begin{equation}
V=\frac{1}{\tau}\int_{0}^{\tau}[\frac{1}{2}m\omega_{0}^{2}x^{2}(t)+\frac{1}{4}\alpha x^{4}(t)]dt,
\end{equation}
with $\tau$ much shorter than $1/\gamma$ but much longer than $1/\omega$, which can be satisfied for {a} system with a large quality factor $Q=\omega/\gamma$. By averaging Eq.~\eqref{energy decay} as $\D \varepsilon
/\D t= -2\gamma E_K$, we obtain the differential equation for $X^2(t)$:
\begin{equation}
\frac{\D X^{2}(t)}{\D t}=-\gamma X^{2}(t) \left [1-\kappa X^{2}(t) +O( \kappa^2  X^4(t))\right ],
\label{dxsqrt}
\end{equation}
Dropping terms of order $\kappa^2  X^4 (t)$ or higher, we obtain the solution of Eq.~\eqref{dxsqrt}
\begin{equation}
X^{2}(t)\left [\kappa-\left( \frac{\kappa X^{2}(0)-1}{X^{2}(0)}\right)e^{\gamma t}\right ] = 1,
\label{xsqrt}
\end{equation}
{Asymptotically, $X(t)$ decays and} Eq.~\eqref{xsqrt} can be expanded as follows:
\begin{eqnarray}\label{eqB10}
X^{2}(t) = X^2(0) e^{-\gamma t}\left [1-\kappa X^2(0)(e^{-\gamma t} -1 )+ ...\right ].
\nonumber\\
\left . \right .
\end{eqnarray}

Next, we  define the autocorrelation function of $X^2(t)$ as:
\begin{equation}
\begin{aligned}
\label{Gxx} G_{X^2}(t) =\langle X^{2}(t)X^{2}(0)\rangle,
\end{aligned}
\end{equation}
which according to Eq.~\eqref{eqB10} becomes
\begin{eqnarray}
G_{X^{2}}(t)=\langle X^{4}(0)\rangle e^{-\gamma t} \left [1-\kappa X^2(0)(e^{-\gamma t} -1 )+ ...\right ].\nonumber\\
\left . \right .
\end{eqnarray}
In {the} experiment, $X^2(t)$ is directly measured from the power spectral density $S_{ x}(\omega) $ by following standard procedures \cite{Poggio2007,Huang2016}, as  $X^2(t) = S_{x}(\omega) b$, where $b$ is the sampling bandwidth
satisfying $\gamma\ll b \ll \omega_0$. We also define the following normalized autocorrelation:
\begin{equation}
\begin{aligned}
\label{R} R_{X^2}(t) =\frac{\langle X^{2}(t)X^{2}(0)\rangle}{\langle X^{4}(0)\rangle},
\end{aligned}
\end{equation}
which is used to estimate the damping rate $\gamma/ 2 \pi$.

\begin{figure}[h!]
\includegraphics[width=0.9\linewidth]{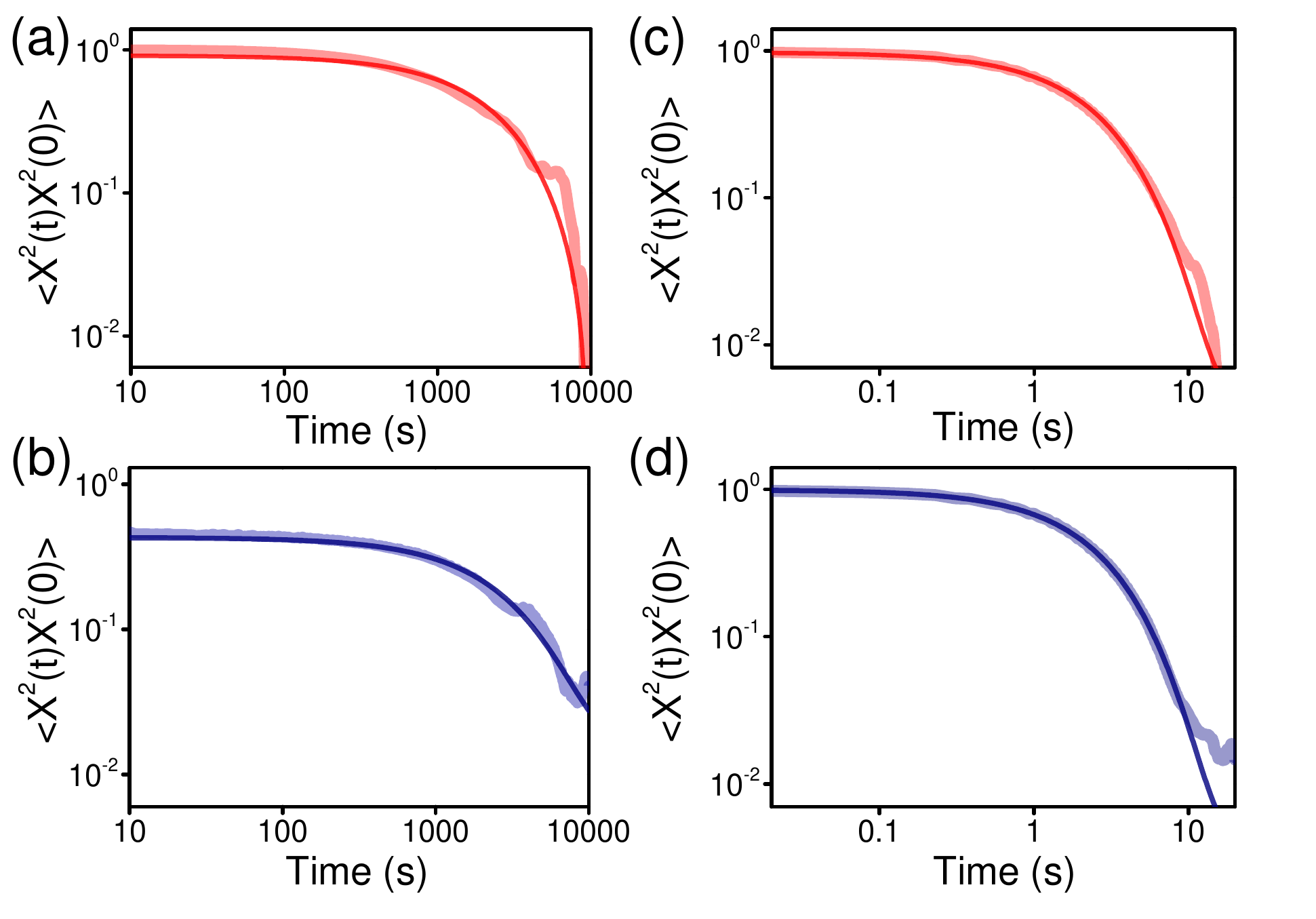}
\caption{Influence of the nonlinearity on the autocorrelation of the oscillator energy defined as $\langle X^2(t) X^2(0) \rangle$ (normalized as per Eq.~\eqref{R}). Panels (a) and (b): numerically simulated autocorrelation (light thick curves)
under high vacuum  (with $\gamma /2\pi=  0.0004\,$Hz) when the nonlinearity is excluded and included, respectively. The curves are fitted with the exponential decay $\exp(-t/\tau)$ (thin curves); the resulting damping rates $\gamma$ turn out to be almost the same as the input ones (see Table~\ref{tab:table2} for the values). Panels (c) and (d): medium vacuum counterparts of (a) and (b), with a larger damping rate  $\gamma /2\pi = 0.4\,$Hz.  The recovered damping rates with and without nonlinearity both well agree well with the input values.}
	\label{decay}
\end{figure}

In our system, nonlinearities {come} not only from the term $\alpha x^3$, but also from the coupling of {the} motion {along} different axes, as $\beta_{i,j} x_{i}x_{j}^2$. {We calculated numerically}
the effects based on two-mode coupling from the equations of motion:
\begin{equation}
\begin{aligned}
\label{EOM4} m  \ddot{x}_1+m \gamma \dot{x}_1  + m \omega_1^2 x_1 + \alpha_1 x_1^3  +  \beta x_2^2 x_1  = f_1(t)  \\
  m  \ddot{x}_2+m \gamma \dot{x}_2  + m \omega_2^2 x_2 +\alpha_2  x_2^3  + \beta x_1^2 x_2  = f_2(t).
\end{aligned}
\end{equation}
Here modes 1 and 2 correspond to {the} motions in {the} $x$-$y$ plane, while the motion along $z$ axis is neglected, $f_1(t)$ and $f_2(t)$ are independent white noise with power spectral density $S_{1,2}$
equal to that of the thermal Brownian noise measured experimentally. The values $m$, $\gamma$ and $\omega_{1,2}$ are directly obtained from {the} experiment.  The nonlinearity coefficients $\alpha_{1,2}$ and $\beta$  are
tuned so that the full width at half maximum  and the shape of the power spectral density $S_{x}(\omega)$ {obtained from the} numerical simulation and from the experiments agree with each other, as shown in  Fig.~\ref{PSD}. The corresponding
values {are} $\alpha_{1} = -6.4\,\rm kg/m^2 s^2$, $\alpha_{2} = -2.1\,\rm kg/m^2 s^2$ and $\beta = 6.4\,\rm kg/m^2 s^2$.  $R_{X^2}(t)$ {is} numerically calculated {for medium and high vacuum} and the results are
shown in Fig.~\ref{decay}. The data are fitted to the exponential decay $\exp(-t/\tau)$, producing the damping rates $\gamma = 1/\tau$, which agree well with the values used in numerical simulations (see
Table~\ref{tab:table2}).

\begin{table}[t]
	\caption{\label{tab:table2} {Comparison of the damping rates.} \textit{input $\gamma$} is the input value of damping rate used in the simulation, \textit{fitted $\gamma$ (nonlinear)} is the result of the simulation with nonlinearity added into the equation of motion, \textit{fitted $\gamma$ (linear)} {that} from the result of simulation without nonlinearity. {The first row corresponds to medium vacuum, while the second row to high vacuum.} }
	\begin{ruledtabular}
		\begin{tabular}{c|ccc}
			&input $\gamma$   & fitted $\gamma$  (nonlinear) & fitted $\gamma$  (linear)     \\ \hline
			$P_\text{MV}$ &0.4\,Hz        &  0.39\,Hz    & 0.38\,Hz   \\ 
			$P_\text{HV}$  &0.0004\,Hz     &      0.00037\,Hz      & 0.00038\,Hz   \\
		\end{tabular}
	\end{ruledtabular}
\end{table}


\section{Error estimation }\label{error}

\begin{figure}[h!]\centering
	\includegraphics[width=1\linewidth]{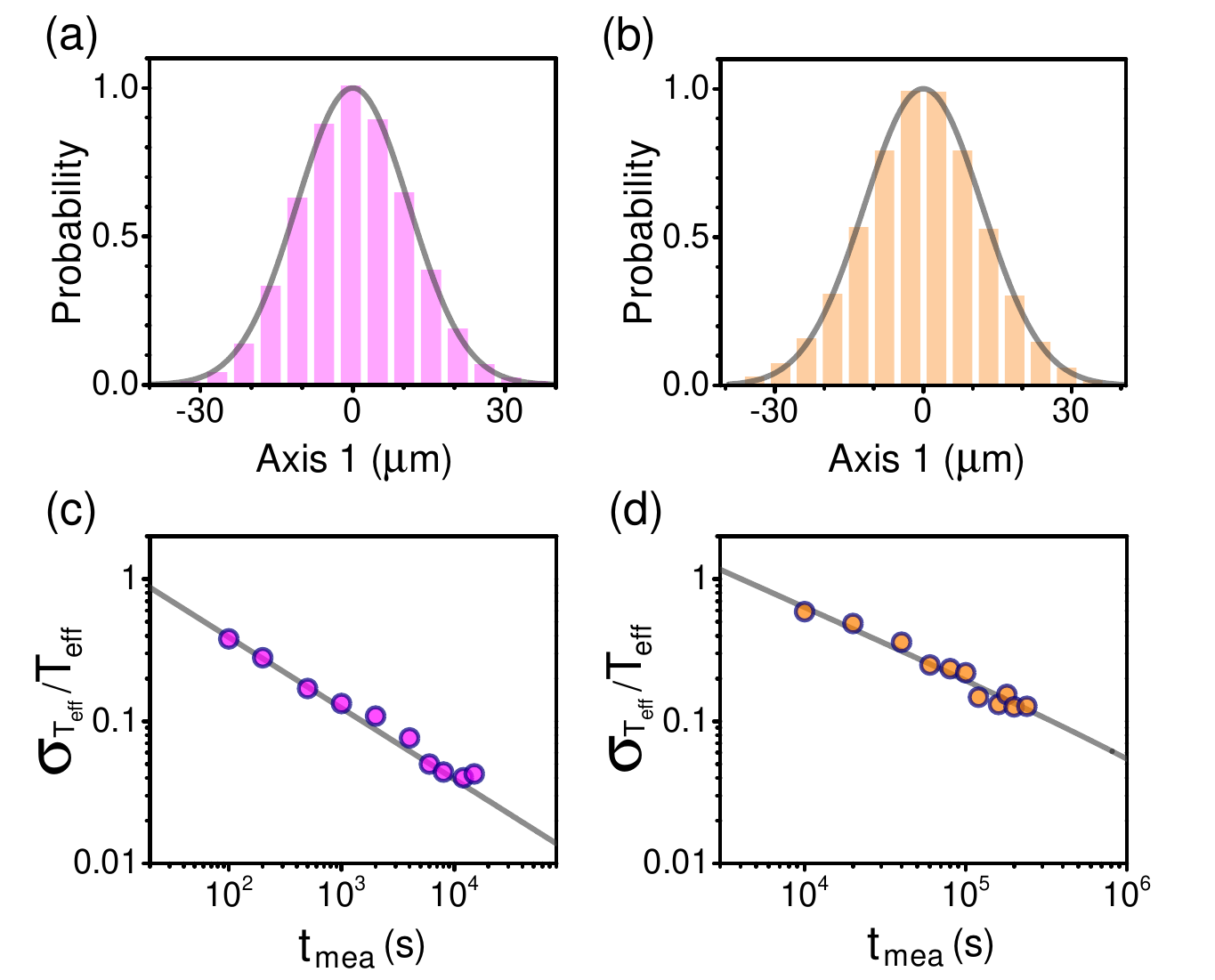}
	\caption{Panel (a): probability distribution of the displacement, measured in medium vacuum ($P \approx 1 \times 10^{\rm -4} \,\rm mbar$), and its fitting with a Gaussian distribution function of standard deviation $\sigma$. Panel (b): same as panel (a) but measured in high vacuum ($P \approx 5\times 10^{\rm -7} \,\rm mbar$). Panel (c):  standard deviation of the relative effective temperature $\sigma_{T_{\rm eff}}(t_{\rm mea})/T_{\rm eff}$ as a function of acquisition time $t_{\rm mea}$ at medium vacuum. The total measurement time is about $7.5\times 10^4$\,s.  The curve shows the theoretical value according to Eq.~\eqref{error decay}. Panel (d): same as panel (c) but measured at high vacuum. The total measurement time is about $9.5\times 10^5$\,s. }
	\label{sigma}
\end{figure}

In order to estimate the error on the effective temperature $T_{\rm eff}$  from the measured  position distribution, the displacement distributions of {the} oscillation mode 1 (12.9\,Hz) under high vacuum (HV) and medium vacuum (MV) were recorded [Fig.~\ref{sigma} (a) and (b)]. The results are fitted to {a} Gaussian distribution to give the standard deviation  $\sigma_{\rm}^{\rm HV}$($\sigma_{\rm}^{\rm MV}$), from which the effective temperatures $T_{\rm eff}^{\rm HV}$($T_{\rm eff}^{\rm MV}$) are obtained. For a given measurement time $t_{\rm mea}$, the standard deviation $\sigma_{T_{\rm eff}}$ of {the} measured effective temperature  can be derived by following the procedure in Refs.~\cite{Ranjit2016} and \cite{Hebestreit2018}.  The results for medium vacuum (MV) and high vacuum (HV) are plotted in  Fig.~\ref{sigma} (c) and (d), respectively,  as functions of $t_{\rm mea}$. Theoretically, the relative standard deviation of {the} effective temperature as a function of the measurement time $t_{\rm mea}$ satisfies the relation \cite{Hebestreit2018}
\begin{equation}
\begin{aligned}
\label{error decay} \frac{\sigma_{T_{\rm eff}}(t_{\rm mea})}{T_{\rm eff}} = \sqrt{\frac{2}{\gamma t_{\rm mea}}},
\end{aligned}
\end{equation}
and is plotted in Fig.~\ref{sigma} (c) and (d) as straight lines. The measured data agree very well with {the} theory.  Finally, the uncertainty $\sigma_{T_{\rm eff}}$ of the effective temperature is estimated using Eq.~\eqref{error decay} by taking $t_{\rm mea} = t_{\rm total}$, the total measurement time. In particular, the total data acquisition time at high vacuum is $9.5\times 10^5$\,s (about 11 days), which can be further extended to reduce {the} uncertainty, but {this was not}  done for practical reasons. The effective temperature measured in a medium vacuum is taken as the environmental temperature $T_{\rm eff}^{\rm MV} = T_{\rm env}$ and the temperature difference is $\delta T = T_{\rm eff}^{\rm HV} - T_{\rm eff}^{\rm MV}$. To estimate the  upper bound on $\delta T$ with the standard methods \cite{Feldman1998}, $T_{\rm eff}^{\rm HV}$ and $ T_{\rm eff}^{\rm MV}$ are treated as independent and both following Gaussian distributions, with their corresponding standard deviations $\sigma_{T_{\rm eff}}$ obtained from {the} measured data [Fig.~\ref{sigma} (c) and (d)]. The threshold $\sigma_{\delta T}$ defined by the $95\%$ confidence level ($\delta T < \sigma_{\delta T}$) is given in Table~\ref{tab:table3}. We note that the measured effective temperature does not coincide with the temperature (298\,K) measured by the thermometer in the vacuum chamber. While such a bias is due to the uncertainty in measuring the absolute displacement of the oscillator, there is an uncertainty of less than a few percent in determining the magnification $\mathcal{M}$ of the detection optics, so is in the micro-sphere's absolute displacement is given by $\mathbf{x} = \mathbf{x^{\prime}}/ \mathcal{M}$. This uncertainty is constant during the whole {experimental} process and only brings about a small error (a few percent) on the final result. Since the power spectral density of additional force noises is defined as $\delta S_{\rm total} = 2 m\gamma k_{\rm B} \delta T $, we estimate its upper bound as $\delta S_{\rm total} < 2 m\gamma k_{\rm B} \sigma_{\delta T}$. Finally, we obtain the upper bounds on the CSL collapse rate $\lambda$ from Eq.~\eqref{eta_CSL3} by using the upper bound on the CSL collapse strength $\eta$ given by $\hbar^{2} \eta < 2 m\gamma k_{\rm B} \sigma_{\delta T}$.

\begin{table}[t!]
	\caption{\label{tab:table3} Measured effective temperature of the oscillator. $T_{\rm eff}^{\rm HV}$ is the effective temperature of the oscillator measured at high vacuum and $T_{\rm eff}^{MV}$ {that} at medium {vacuum}.  $T_{\rm eff}^{MV}= T_{\rm env}$ is used for the estimation of the environment temperature. $\delta T$ is defined as the temperature increase $\delta T = T_{\rm eff}^{HV} - T_{\rm env}$. {$\sigma_{ T_{\rm eff}^{HV}}$,} $\sigma_{ T_{\rm eff}^{MV}}$ and $\sigma_{\delta T}$ are the corresponding upper bounds at $95\%$ confidence level. 
 }
	\begin{ruledtabular}
		\begin{tabular}{cc|cc|cc}
		\multicolumn{2}{c}{High Vacuum}&\multicolumn{2}{c}{Medium Vacuum}&\multicolumn{2}{c}{Difference}\\
		\hline
			$ T_{\rm eff}^{\rm HV}$   & $\sigma_{T_{\rm eff}^{\rm HV}}$  &   $ T_{\rm eff}^{\rm MV}$  &  $\sigma_{T_{\rm eff}^{\rm MV}}$ & $\delta T $ &$\sigma_{ \delta T }$ \\
   \hline
   			297.9\,K    & 16.2\,K  & 291.4\,K   &  4.1\,K & 6.5\,K &      40\,K     \\

		\end{tabular}
	\end{ruledtabular}

\end{table}

\begin{thebibliography}{99}


\bibitem{Bohm1} D. Bohm, \textit{A Suggested Interpretation of the Quantum Theory in Terms of "Hidden" Variables. I}, Phys. Rev. \textbf{85}, 166 (1952).

\bibitem{Bohm2} D. Bohm, \textit{A Suggested Interpretation of the Quantum Theory in Terms of "Hidden" Variables. II}, Phys. Rev. \textbf{85}, 180 (1952).

\bibitem{Durr} D. D\"urr and S. Teufel, \textit{Bohmian Mechanics}, Springer-Verlag Berlin Heidelberg (2009).

\bibitem{manyworlds1} H. Everett, \textit{Theory of the Universal Wavefunction}, Thesis, Princeton University, (1956).

\bibitem{manyworlds2} H. Everett, \textit{"Relative State" Formulation of Quantum Mechanics}, Rev. Mod. Phys. \textbf{29}, 454 (1957).

\bibitem{wallace} D. Wallace, \textit{Everettian Rationality: defending Deutsch's approach to probability in the Everett interpretation}, Stud. Hist. Phil. Mod. Phys. \textbf{34}, 415 (2003).

\bibitem{dariano} G. Chiribella, G.M. D?Ariano, P. Perinotti, \textit{Informational derivation of quantum theory}, Phys. Rev. A \textbf{84}, 012311 (2011).

\bibitem{GRW} G.C. Ghirardi, A. Rimini, and T. Weber, \textit{Unified dynamics for microscopic and macroscopic systems}, Phys. Rev. D \textbf{34}, 470 (1986).

\bibitem{Pearle1989} Philip Pearle, \textit{Combining stochastic dynamical state-vector reduction with spontaneous localization}, Phys. Rev. A \textbf{39}, 2277 (1989).

\bibitem{Ghirardi1990} G.C. Ghirardi, P. Pearle, and A. Rimini, \textit{Markov processes in Hilbert space and continuous spontaneous localization of systems of identical particles}, Phys. Rev. A \textbf{42}, 78 (1990).

\bibitem{Bassi2003} A. Bassi, G.  Ghirardi, \textit{Dynamical reduction models}, Phys. Rep. \textbf{379}, 257 (2003).

\bibitem{Bassi2013} A. Bassi, K. Lochan, S. Satin, T. Singh, and H. Ulbricht,  \textit{Models of wave-function collapse, underlying theories, and experimental tests}, Rev. Mod. Phys. \textbf{85}, 471 (2013).

\bibitem{Nimmrichter2014}  S. Nimmrichter, K. Hornberger, and K. Hammerer, \textit{Optomechanical Sensing of Spontaneous Wave-Function Collapse},  Phys. Rev. Lett. \textbf{113}, 020405 (2014).

\bibitem{bahrami2014} M. Bahrami, M. Paternostro, A. Bassi, and H. Ulbricht, \textit{Proposal for a Noninterferometric Test of Collapse Models in Optomechanical Systems}, Phys. Rev. Lett. \textbf{112}, 210404 (2014).

\bibitem{Diosi2014} L.Di\'{o}si.  \textit{Testing spontaneous wave-function collapse models on classical mechanical oscillators},  Physical Review Letters, \textbf{114},050403 (2015).

\bibitem{Leggett} A. J. Leggett, \textit{Testing the limits of quantum mechanics: motivation, state of play, prospects}, J. Phys.: Condens. Matter, \textbf{14},  R415 (2002).

\bibitem{Diosi1989} L. Di\'{o}si,\textit{Models for universal reduction of macroscopic quantum fluctuations},  Phys. Rev. A \textbf{40}, 1165 (1989)

\bibitem{Penrose1996} R. Penrose, \textit{On Gravity's role in Quantum State Reduction}, Gen. Relativ. Gravit. \textbf{28}, 581 (1996)

\bibitem{Penrose} R. Penrose, \textit{The Road to Reality: A Complete Guide to the Laws of the Universe}, Jonathan Cape (2004).

\bibitem{Adler2007} S. L. Adler, \textit{Lower and upper bounds on CSL parameters from latent image formation and IGM~heating}, J. Phys. A: Math. Theor. \textbf{40}, 2935 (2007).

\bibitem{Kovachy2015} T. Kovachy, P. Asenbaum, C. Overstreet, C. A. Donnelly, S. M. Dickerson, A. Sugarbaker, J. M. Hogan and M. A. Kasevich, \textit{Quantum superposition at the half-metre scale}, Nature volume \textbf{528},  530-533 (2015).

 \bibitem{Coldmolecular1999} M. Arndt, O. Nairz, J. Vos-Andreae, C. Keller, G. van der Zouw and A. Zeilinger, \textit{Wave¨Cparticle duality of C60 molecules}, Nature \textbf{28} \textbf{401}, 680 (1999).

\bibitem{Coldmolecular2011} S. Gerlich, S. Eibenberger, M. Tomandl, S. Nimmrichter, K. Hornberger, P. J. Fagan, J. T¨¹xen, M. Mayor and M. Arndt, \textit{Quantum interference of large organic molecules}, Nat. Commun.
    \textbf{2}, 263 (2011).

\bibitem{Coldmolecular2012} K. Hornberger, S. Gerlich, P. Haslinger, S. Nimmrichter and M. Arndt, Rev. Mod. Phys. \textit{Quantum interference of clusters and molecules}, \textbf{84}, 157 (2012).

\bibitem{Eibenberger2013} S. Eibenberger, S. Gerlich, M. Arndt, M. Mayor and J. T\"uxen, \textit{Matter?wave interference of particles selected from a molecular library with masses exceeding 10\,000 amu}, Phys. Chem. Chem. Phys. \textbf{15}, 14696-14700 (2013).

\bibitem{Interf1} M.Toro\v s, G.Gasbarri, A.Bassi , \textit{ Colored and dissipative continuous spontaneous localization model and bounds from matter-wave interferometry}, Phys. Rev. A, \textbf{381} 3921-3927 (2017).

 \bibitem{Interf2}  M.Toro\v s, A.Bassi . \textit{ Bounds on quantum collapse models from matter-wave interferometry: calculational details}, Journal of Physics A: Mathematical and Theoretical, \textbf{51} 115302 (2018).

\bibitem{Diamonds}S.Belli , R.Bonsignori , G.D'Auria , et al., \textit{ Entangling macroscopic diamonds at room temperature: Bounds on the continuous-spontaneous-localization parameters}, Phys. Rev. A, \textbf{94}, 012108 (2016).

\bibitem{Diamondsexp} KC. Lee \textit{et al.}, \textit{Entangling macroscopic diamonds at room temperature}, Science \textbf{334}, 1253 (2011).

\bibitem{ultra-cold atom} T. Kovachy, J. M. Hogan, A. Sugarbaker, S. M. Dickerson, C. A. Donnelly, C. Overstreet, and M. A. Kasevich, Phys. Rev. Lett.  \textbf{114}, 143004 (2015)

\bibitem{ultra-cold atom2} M. Bilardello, S. Donadi, A. Vinante, A. Bassi, \textit{Bounds on collapse models from cold-atom experiments}, Physica A \textbf{462}, 764 (2016)

\bibitem{residual_heating} E. Nazaretski, V. O. Kostroun, S. Dimov, R. O. Pohl, and J. M. Parpia, \textit{Heat Inputs to Sub-mK Temperature Cryostats and Experiments from $\gamma$- Radiation and Cosmic Ray Muons}, J. Low
    Temp. Phys. \textbf{137}, 609 (2004).

\bibitem{Adler2018} S. L.  Adler and A. Vinante, \textit{Bulk heating effects as tests for collapse models}, Phys. Rev. A \textbf{97},  052119 (2018).

\bibitem{Phonos2} M. Bahrami, \textit{Testing collapse models by a thermometer},Phys. Rev. A.\textbf{97},052118 (2018).

\bibitem{Planetary} I. de Pater and J. J. Lissauer, \textit{Planetary Sciences}, Cambridge University Press, Cambridge, (2001).

\bibitem{Planets} S.L. Adler, A. Bassi, M. Carlesso and A. Vinante, \textit{Testing Continuous Spontaneous Localization with Fermi liquids}, Phys. Rev. D.\textbf{99},103001 (2019).

\bibitem{Planets2} A. Tilloy and TM. Stace, \textit{Neutron Star Heating Constraints on Wave-Function Collapse Models}, Phys. Rev. Lett. \textbf{123}, 080402 (2019).

\bibitem{x-raycase} A. Bassi and S. Donadi, \textit{Spontaneous photon emission from a non-relativistic free charged particle in collapse models: A case study}, Physics Letters A \textbf{378}, 761-765 (2014).


\bibitem{X-ray} C. Curceanu, B. C. Hiesmayr, K. Piscicchia, \textit{X-rays help to unfuzzy the concept of measurement}, J. Adv. Phys. \textbf{4}, 263 (2015).

\bibitem{X-ray2} K.Piscicchia , A.Bassi , C.Curceanu , et al., \textit{CSL Collapse Model Mapped with the Spontaneous Radiation}, Entropy, \textbf{19},319-(2017).

\bibitem{AURIGA} A. Vinante et al., (AURIGA Collaboration),\textit{Present performance and future upgrades of the AURIGA capacitive readout Classical Quantum Gravity} \textbf{23}, S103 (2006).

\bibitem{LIGO1} B.P. Abbott et al. (LIGO Scientific Collaboration and Virgo Collaboration), \textit{GW150914: The Advanced LIGO Detectors in the Era of First Discoveries},Phys. Rev. Lett. \textbf{116}, 131103 (2016).

\bibitem{LISA1} M. Armano et al., \textit{Sub-Femto-g Free Fall for Space-Based Gravitational Wave Observatories: LISA Pathfinder Results}, Phys. Rev. Lett. \textbf{116}, 231101 (2016).

\bibitem{LISA2} M. Armano et al.,  \textit{Beyond the Required LISA Free-Fall Performance: New LISA Pathfinder Results down to 20$\mu$Hz}, Phys. Rev. Lett. \textbf{120}, 061101 (2018).

\bibitem{gravitation results1}    M.Carlesso, A.Bassi, P.Falferi and A.Vinante, \textit{Experimental bounds on collapse models from gravitational wave detectors} Phys. Rev. D \textbf{94}, 124036 (2016).

\bibitem{gravitation results2} B. Helou, B.J.J. Slagmolen, D. E. McClelland, and Y. Chen, \textit{LISA pathfinder appreciably constrains collapse models}, Phys. Rev. D \textbf{95}, 084054 (2017).

 \bibitem{Rotational1}   M.Carlesso,M.Paternostro,H.Ulbricht, et al., \textit{Non-interferometric test of the Continuous Spontaneous Localization model based on the torsional motion of a cylinder},  New J. Phys. \textbf{20} 0830222017 (2018).

\bibitem{Vinante2016} A. Vinante, M. Bahrami, A. Bassi, O. Usenko, G. Wijts, and T.H. Oosterkamp, \textit{Upper Bounds on Spontaneous Wave-Function Collapse Models Using Millikelvin-Cooled Nanocantilevers}, Phys. Rev. Lett.
    \textbf{116}, 090402 (2016).

\bibitem{Vinante2017} A. Vinante, R. Mezzena, P. Falferi, M. Carlesso, and A. Bassi, \textit{Improved Noninterferometric Test of Collapse Models Using Ultracold Cantilevers}, Phys. Rev. Lett. \textbf{119}, 110401 (2017).

\bibitem{Pontin2019} A. Pontin, N. P. Bullier, M. Toro\v s, P. F. Barker, \textit{An ultra-narrow line width levitated nano-oscillator for testing dissipative wavefunction collapse}, ArXiv: 1907.06046 (2019).

\bibitem{Collett2003} B. Collett and P. Pearle, \textit{Wavefunction Collapse and Random Walk}, Found. Phys. \textbf{33},1495 (2003).

\bibitem{levitation3} D. Goldwater, M. Paternostro, and P.F. Barker, \textit{Testing wave-function-collapse models using parametric heating of a trapped nanosphere}, Phys.Rev. A \textbf{94}, 010104 (2016).

\bibitem{MAQRO} R. Kaltenbaek \textit{et al.}, \textit{Macroscopic Quantum Resonators (MAQRO): 2015 update}, Eur. Phys. J. Quant. Tech. \textbf{3}, 5 (2016).

\bibitem{McMillen2017} S. McMillen, M. Brunelli, M. Carlesso, A. Bassi, H. Ulbricht, M. G. A. Paris, and M. Paternostro, \textit{Quantum-limited estimation of continuous spontaneous localization}, Phys. Rev. A \textbf{95}, 012132 (2017).

\bibitem{rotschrinski} B. Schrinski, B. A. Stickler, and K. Hornberger, \textit{Collapse-induced orientational localization of rigid rotors}, J. Opt. Soc. Am. B \textbf{34}, C1 (2017).

\bibitem{multilayer} M. Carlesso, A. Vinante, and A. Bassi, \textit{Multilayer test masses to enhance the collapse noise}, Phys. Rev. A \textbf{98}, 022122 (2018).

\bibitem{mishra2018} R. Mishra, A. Vinante, and T. P. Singh, \textit{Testing spontaneous collapse through bulk heating experiments: An estimate of the background noise}, Phys. Rev. A \textbf{98}, 052121 (2018).

\bibitem{colored1} A. Bassi, L. Ferialdi, \textit{Non-Markovian dynamics for a free quantum particle subject to spontaneous collapse in space: general solution and main properties}, Phys. Rev. A \textbf{80}, 012116 (2009).

\bibitem{colored2} L. Ferialdi, A. Bassi, \textit{Dissipative collapse models with nonwhite noises} Phys. Rev. A \textbf{86}, 022108 (2012).

\bibitem{colored3} S.L. Adler, A. Bassi, S. Donadi, \textit{On spontaneous photon emission in collapse models}, J. Phys. A \textbf{46}, 245304 (2013).

\bibitem{Colored-opto} M. Carlesso, L. Ferialdi and A. Bassi, \textit{Colored collapse models from the non-interferometric perspective}, Eur. Phys. J. D \textbf{72}, 159 (2018).

\bibitem{opto-rev} M. Aspelmeyer, T.J. Kippenberg, F. Marquardt, \textit{Cavity Optomechanics}, Rev. Mod. Phys. \textbf{86}, 1391 (2014).

\bibitem{gravitation results3} A. Vinante, A. Pontin, M. Rashid, M. Toros, P.F. Barker, and H. Ulbricht, \textit{Testing collapse models with levitated nanoparticles: the detection challenge},Phys. Rev. A \textbf{100}.012119 (2019).

\bibitem{levitation1} O. Romero-Isart, \textit{Quantum superposition of massive objects and collapse models}, Phys. Rev. A \textbf{84}, 052121 (2011)

\bibitem{levitation2} Z. Yin, T. Li, X. Zhang  and L. M. Duan, \textit{Large quantum superpositions of a levitated nanodiamond through spin-optomechanical coupling}, Phys. Rev. A \textbf{88}, 033614 (2013).

\bibitem{levitation4}    J. Li, S. Zippilli, J. Zhang, and D. Vitali,\textit{Discriminating the effects of collapse models from environmental diffusion with levitated nanospheres}, Phys. Rev. A \textbf{93}, 050102(R)
    (2016).

\bibitem{rugar1995}  J. A. Sidles, J. L. Garbini, K. J. Bruland, D. Rugar, O. Z\"{u}ger, S. Hoen, and C. S. Yannoni, \textit{Magnetic resonance force microscopy}, Rev. Mod. Phys. \textbf{67}, 249 (1995).

\bibitem{Clerk2010} A. A. Clerk, M. H. Devoret, S. M. Girvin, F. Marquardt, and R. J. Schoelkopf, \textit{Introduction to quantum noise, measurement, and amplification},  Rev. Mod. Phys. \textbf{82}, 1155 (2010).

\bibitem{huangpu1}  P. Huang, et al.,\textit{Demonstration of motion transduction based on parametrically coupled mechanical Resonators} Phys. Rev. Lett.  \textbf{110}, 227202 (2013).

\bibitem{rugar2004} D. Rugar, R. Budakian, H. J. Mamin and B. W. Chui, \textit{Single spin detection by magnetic resonance force microscopy}, Nature \textbf{430}, 329 (2004).

\bibitem{Kippenberg2012} E. Gavartin, P. Verlot and T. J. Kippenberg, \textit{A hybrid on-chip optomechanical transducer for ultrasensitive force measurements},Nat. Nanotech. \textbf{7},  509 (2012).

\bibitem{Geim1999} A. K. Geim, M. D. Simon, M. I. Boamfa and L. O. Heflinger, \textit{Magnet levitation at your fingertips}, Nature, \textbf{400}, 323 (1999).

\bibitem{Slezak2018} B. R. Slezak, C. W. Lewandowski, J. Hsu, and Brian D'Urso, \textit{Cooling the motion of a silica microsphere in a magneto-gravitational trap in ultra-high vacuum},\textbf{20}, 063028 (2018).

\bibitem{optical_levitaion1} D. E. Chang, C. A. Regal, S. B. Pappb,  D. J. Wilsonb£¬J. Yeb, O. Painterd, H. J. Kimble, and P. Zoller, \textit{ Cavity opto-mechanics using an optically levitated nanosphere}, Proc. Nat. Acad.
    Sci.\textbf{107}, 1005 (2010)

\bibitem{Dykman1992} M. I. Dykman, P. V. E. McClintock, \textit{ Power spectra of noise-driven nonlinear systems and stochastic resonance}. Physica D \textbf{58} (1-4), 10¨C30 (1992).

\bibitem{rugar2001} B. C. Stipe, H. J. Mamin, T. D. Stowe, T. W. Kenny, and D. Rugar, \textit{ Noncontact Friction and Force Fluctuations between Closely Spaced Bodies}, Phys. Rev. Lett. \textbf{120}, 096801 (2001).

\bibitem{optical_levitaion3} J. Gieseler, L. Novotny and R. Quidant, \textit{ Thermal nonlinearities in a nanomechanical oscillator}, Nat. Phys. \textbf{9}, 806 (2013).

\bibitem{electric_levitation} J. Millen, P. Z. G. Fonseca, T. Mavrogordatos, T. S. Monteiro, and P. F. Barker, \textit{Cavity Cooling a Single Charged Levitated Nanosphere}, Phys. Rev. Lett. \textbf{114}, 123602 (2015).

\bibitem{optical_levitaion2} T. Li, S. Kheifets,  and M. G. Raizen, \textit{ Millikelvin cooling of an optically trapped microsphere in vacuum}, Nat. Phys. \textbf{7}, 527 (2011).

\bibitem{Pino} H. Pino, J. Prat-Camps, K. Sinha, B. Prasanna Venkatesh and O. Romero-Isart, \textit{On-chip quantum interference of a superconducting microsphere}, Quantum Sci. Technol. \textbf{3}, 25001 (2018).


\bibitem{Magnet205} Romero-Isart O , Clemente L , Navau C , et al. \textit{ Quantum Magnetomechanics with Levitating Superconducting Microspheres},Phys. Rev. Lett. \textbf{109}, 147205 (2012) .

 \bibitem{Magnet206} M. Cirio, G.K. Brennen, and J. Twamley, \textit{ Quantum Magnetomechanics: Ultrahigh-Q-Levitated Mechanical Oscillators}
 Phys. Rev. Lett. \textbf{109}, 147206 (2012) .


\bibitem{Nonlinear2009} R.Lifshitz, M. C.Cross, Review of Nonlinear Dynamics and Complexity (Wiley-VCH, 2009)

\bibitem{parametric cool} J. Gieseler, B. Deutsch, R. Quidant, and L. Novotny, \textit{ Subkelvin Parametric Feedback Cooling of a Laser-Trapped Nanoparticle},  Phys. Rev. Lett. \textbf{109}, 103603 (2012).

\bibitem{Kampen1976} N. G. van Kampen, \textit{Stochastic differential equations}, Physics Reports \textbf{24}, 171 (1976).

\bibitem{Berry1997} M. V. Berry and A. K. Geim, \textit{Of flying frogs and levitrons},Eur. J. Phys. \textbf{18}, 307 (1997)

\bibitem{Poggio2007}    M. Poggio, C. L. Degen, H. J. Mamin, and D. Rugar,  \textit{ Feedback Cooling of a Cantilever¡¯s Fundamental Mode below 5 mK},  Phys. Rev. Lett. \textbf{99}, 017201 (2007).

\bibitem{Huang2016} P. Huang, J.Zhou, L. Zhang, D. Hou, S. Lin, W.Deng, C. Meng, C. Duan, C. Ju, X. Zheng, Fei Xue, and  J. Du,  \textit{Generating giant and tunable nonlinearity in a macroscopic mechanical resonator from a
    single chemical bond}, Nat. Commun. \textbf{7}, 11517 (2016).

\bibitem{nonlinear1980} M.Dykman, M.Krivoglaz, Fluctuations in nonlinear systems near bifurcations corresponding to the appearance of new stable states. Physica A 104, 480¨C494 (1980)

\bibitem{Ranjit2016} G. Ranjit, M. Cunningham, K. Casey, and A. A. Geraci,  \textit{Zeptonewton force sensing with nanospheres in an optical lattice}, Phys. Rev. A \textbf{93}, 053801 (2016).
\bibitem{Hebestreit2018} E. Hebestreit, M. Frimmer, R. Reimann, C. Dellago, F. Ricci, and L. Novotny, \textit{Calibration and energy measurement of optically levitated nanoparticle sensors}, Rev. Sci. Inst. \textbf{7}, 11517 (2016).\textbf{89}, 033111 (2018).

\bibitem{Feldman1998} G. J. Feldman and R. D. Cousins, \textit{Unified approach to the classical statistical analysis of small signals}, Phys. Rev. D \textbf{ 57}, 3873 (1998)

\end{thebibliography}
\end{document}